\pdfoutput=1	
\documentclass[9pt,column,twoside,conference]{IEEEtran}

\usepackage{framed}
\usepackage{moreverb}
\usepackage[small,sf,bf]{subfigure}
\usepackage{xspace}
\usepackage{flushend}			%

\usepackage{xfrac}			%

\usepackage{amsfonts}
\usepackage{amssymb}    
\usepackage{amsmath}
\usepackage[thmmarks,amsmath,hyperref]{ntheorem}
\usepackage{graphicx}
\usepackage{times}
\usepackage[scaled]{helvet} 		%
\usepackage{textcomp}
\usepackage{color}
\usepackage{colortbl}
\usepackage{booktabs}
\usepackage{listings}
\usepackage{multicol}
\usepackage{times}
\usepackage{pifont}			%
\usepackage{microtype}

\usepackage{mmm}
\usepackage{wand-sample}
\usepackage[scaled=0.85]{beramono}
\usepackage[T1]{fontenc}
\usepackage{url}
\usepackage{multirow}
\usepackage{array}

\definecolor{listinggreen}{rgb}{0,0.6,0}
\definecolor{listinggray}{rgb}{0.5,0.5,0.5}
\definecolor{listingmauve}{rgb}{0.58,0,0.82}
\definecolor{listingkeywordcolor}{rgb}{1.0,0.4,0.0}
\definecolor{listinglightgray}{rgb}{0.9863,0.9863,0.9863}

\lstdefinelanguage{FSharp}
{
	morekeywords	= {
    let,
    type,
    Measure,
	},
	sensitive	= false,
	morecomment	= [l]{\#},
	morecomment	= [s]{(*}{*)},
}

\lstdefinelanguage{Newton}
{
	morekeywords	= {
		signal,
		derivation,
		symbol,
		name,
		invariant,
		constant,
		English,
		sensor,
		name,
		none,
		dot,
		cross,
		derivative,
		integral,
		interface,
		i2c,
		spi,
		analog,
		write,
		read,
		delay,
		range,
		erasuretoken,
		uncertainty,
		accuracy,
		precision,
		Gaussian,
		exponential,
		biexponential,
		to,
		bits,
		dimensionless,
		include
	},
	sensitive	= false,
	morecomment	= [l]{\#},
	morecomment	= [s]{/*}{*/},
}

\usepackage{hyperref}
\hypersetup{
    colorlinks=false, %
    linktoc=all,     %
    linkcolor=blue,  %
}

\usepackage[bordercolor=shadeOne, backgroundcolor=shadeTwo, linecolor=red, textwidth=0.8in, textsize=tiny]{todonotes}
\usepackage{ulem}

\definecolor{shadeOne}{rgb}{0.9,0.95,0.95}
\definecolor{shadeTwo}{rgb}{0.99,0.99,0.99}

\begin{document}
	
	\title{GFET Lab: A Graphene Field-Effect Transistor TCAD Tool}
	
	\author{\IEEEauthorblockN{Nathaniel J. Tye\\}
		\IEEEauthorblockA{\textit{Cambridge Graphene Centre,} \\
			\textit{Department of Engineering,} \\ 
			\textit{University of Cambridge}\\
			njt48@cam.ac.uk \\}
		\and
		\IEEEauthorblockN{Abdul Wadood Tadbier \\}
		\IEEEauthorblockA{\textit{Department of Engineering} \\
			\textit{University of Cambridge}\\
			awt34l@cam.ac.uk}
		\and
		\IEEEauthorblockN{Stephan Hofmann \\}
		\IEEEauthorblockA{\textit{Department of Engineering} \\
			\textit{University of Cambridge}\\
			sh315@cam.ac.uk}
		\and
		\IEEEauthorblockN{Phillip Stanley-Marbell \\}
		\IEEEauthorblockA{\textit{Department of Engineering} \\
			\textit{University of Cambridge}\\
			phillip.stanley-marbell@eng.cam.ac.uk}
	}

	\vspace{-0.1in}
	
	
	\maketitle
	
	%
%
\begin{abstract}
Graphene field-effect transistors (GFETs) are experimental devices which are increasingly seeing commercial and research applications. Simulation and modelling forms an important stage in facilitating this transition, however the majority of GFET modelling relies on user implementation. To this end, we present GFET Lab, a user-friendly, open-source software tool for simulating GFETs.

We first provide an overview of approaches to device modelling and a brief survey of GFET compact models and limitations. From this survey, we identify three key criteria for a suitable predictive model for circuit design: it must be a compact model; it must be SPICE-compatible; it must have a minimal number of fitting parameters. We selected Jimenez's drain-current model as it best matched these criteria, and we introduce some modifications to improve the predictive properties, namely accounting for saturation velocity and the asymmetry in n- and p-type carrier mobilities.

We then validate the model by comparing GFETs simulated in our tool against experimentally-obtained GFET characteristics with the same materials and geometries and find good agreement between GFET Lab and experiment. We demonstrate the ability to export SPICE models for use in higher level circuit simulations and compare SPICE simulations of GFETs against GFETs simulated in GFET Lab, again showing good agreement.

Lastly, we provide a brief tutorial of GFET Lab to demonstrate and encourage its use as a community-developed piece of software with both research and educational applications.
\end{abstract}

	%
%

\section{Introduction}
\label{section:gfet-tool}
Since the observation of the electric field effect in atomically thin 
carbon by Geim and Novoselov in 2004~\cite{Novoselov2004}, there has been a proliferation of 
research into graphene and related materials (GRMs). A key focus of research 
into electronic devices based on GRMs is the realisation of field-effect transistors
(FETs), due to their potential in overcoming the limitations of conventional CMOS 
devices. One of the main attractions of GRMs is that they overcome short channel effects. 

Graphene FETs (GFETs) have a channel made of single- or multi-layer graphene, 
rather than a semiconducting material such as silicon or germanium~\cite{Schwierz2010}.
Unlike traditional semiconductors, graphene's conduction and valence bands don’t
overlap and instead meet at
a single point, known as the \textit{Dirac point}.
Electrons at the Dirac point are effectively massless and so can have very high electron 
mobilities, with the room-temperature phonon-limited
carrier mobility of graphene on SiO\textsubscript{2} being predicted to be around 200,000\,$\mbox{cm}^2\mbox{V}^{-1}\mbox{s}^{-1}$~\cite{Chen2008}.

\begin{figure}[h]
	\centering
	\subfigure[]{\includegraphics[trim=0cm 13cm 23cm 0cm, clip=true, width=0.49\linewidth]{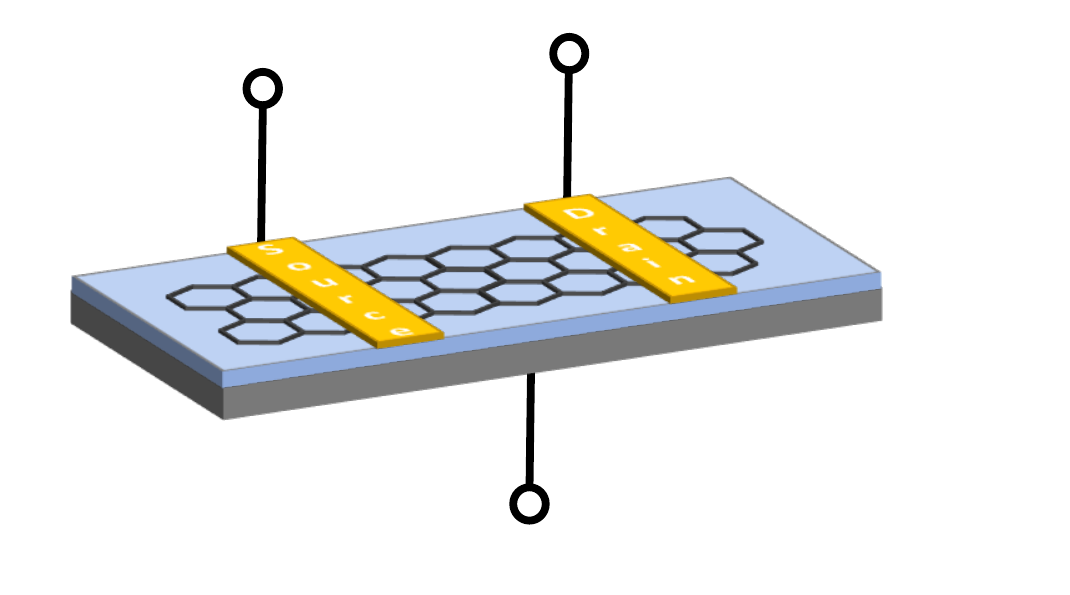}}
	\subfigure[]{\includegraphics[trim=0cm 13cm 23cm 0cm, clip=true, width=0.49\linewidth]{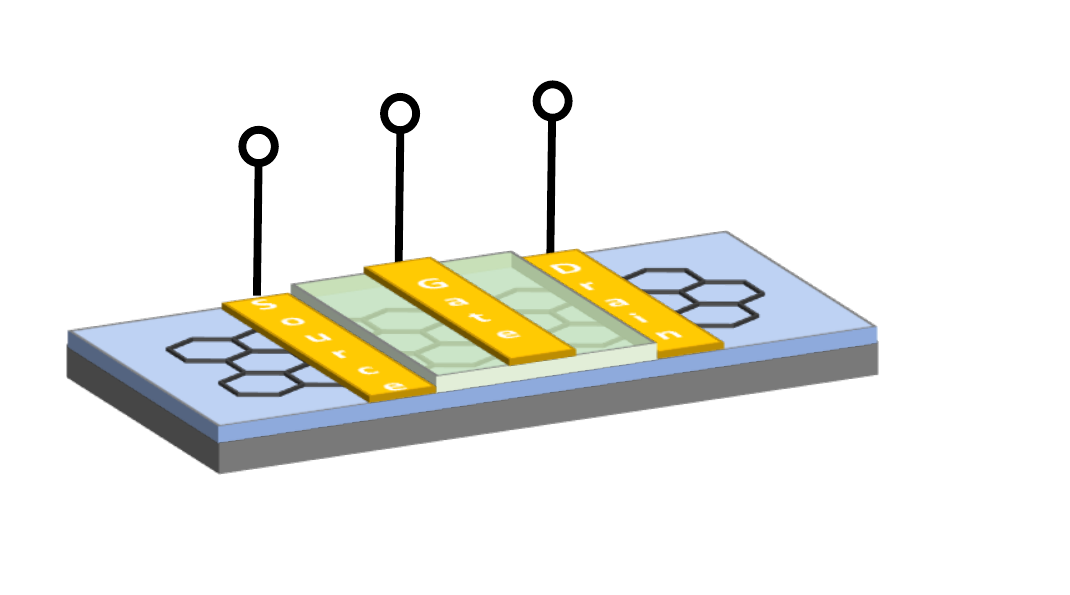}}
	\subfigure[]{\includegraphics[trim=0cm 13cm 23cm 0cm, clip=true, width=0.49\linewidth]{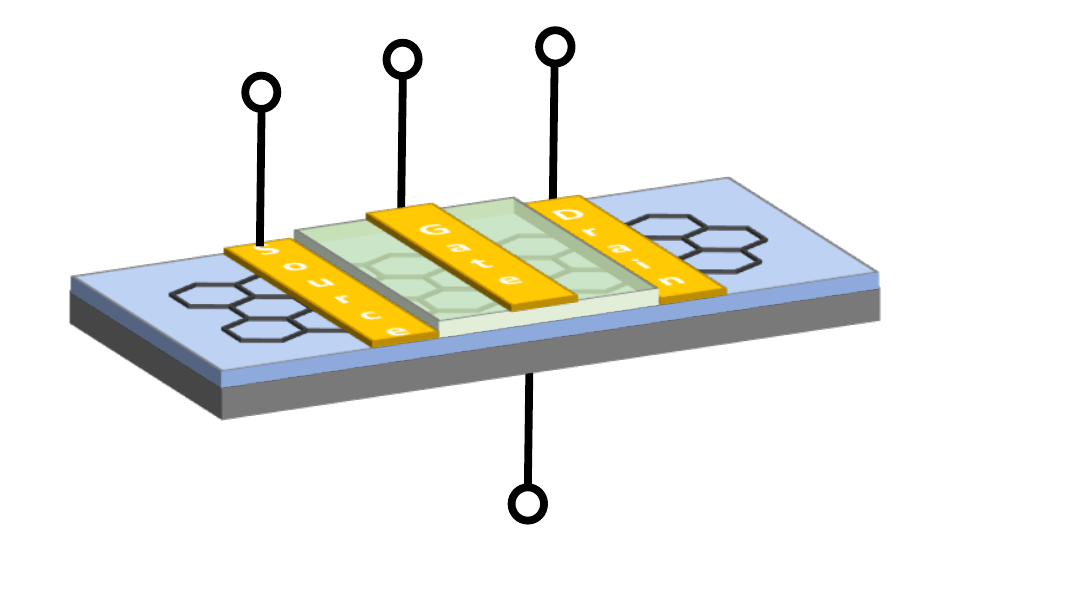}}
	\subfigure[]{\includegraphics[trim=0cm 3cm 5cm 0cm, clip=true, width=0.45\linewidth]{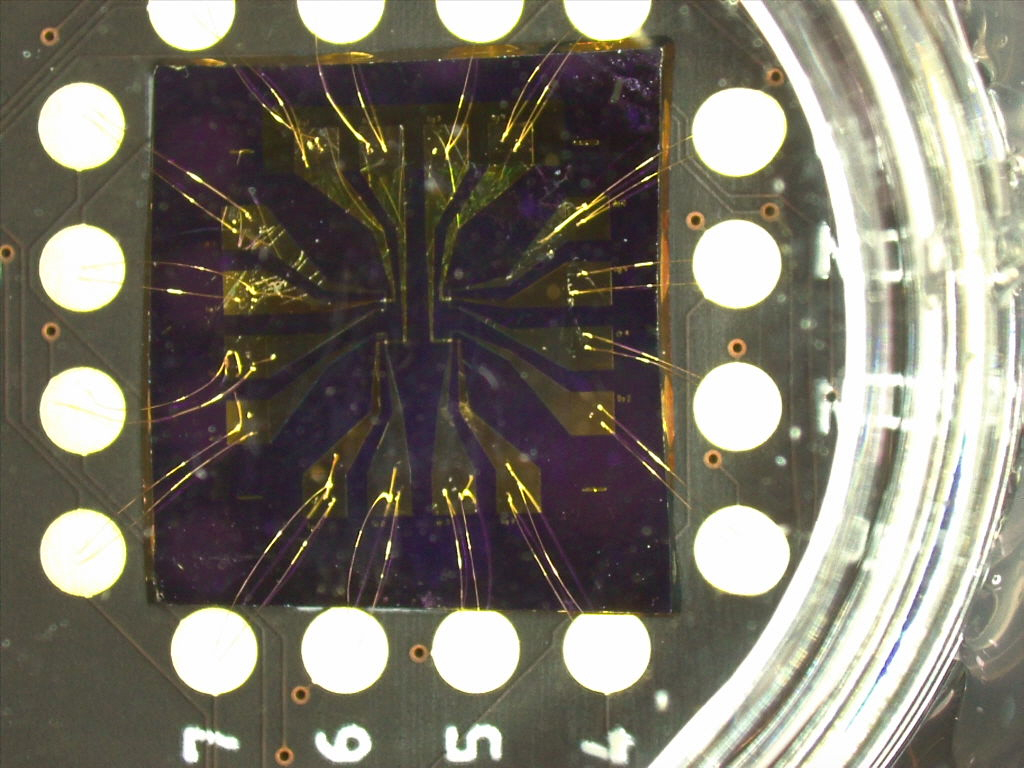}}
	\caption{Illustrations of the three most common forms of GFET. \textbf{(a)} is a back-gated GFET, \textbf{(b)} is a top-gated GFET, and \textbf{(c)} is a dual-gated GFET. The dark grey layer at the bottom represents a doped semiconducting substrate (e.g., p-type Si), the purple layer represents the oxide layer (e.g., SiO\textsubscript{2}), the graphene channel is represented by the hexagonal layer, the top dielectric (e.g., AlO\textsubscript{2}, HfO\textsubscript{2}...) is represented by the light green layer and the relevant contacts are as illustrated. The circuit terminals are also shown in the figures. \textbf{(d)} shows the GFET die we fabricated and characterised in~\cite{Tye2020} and which we use to validate our simulations in Figure~\ref{figure:ToolVsData}\textbf{(a)} and \textbf{b}.}
	\label{figure:GFETSchems}
\end{figure}

Although high electron mobilities result in more efficient flow of charge,
the lack of a band gap means GFETs have low on- to off-current ratios
and can never completely turn off, making them unsuitable for
digital logic applications~\cite{Keyes1985}. However, the poor on- to off-current 
ratios of GFETs in digital logic applications does not preclude their use 
in other areas. GFET-based sensors have applications
in gas~\cite{Schedin2007}, biological~\cite{Lerner2014}, and chemical~\cite{Lu2010} 
sensing applications, and wafers consisting of multiple GFETs are available 
off-the-rack from companies including Graphenea~\cite{Graphenea} and 
BGT Materials~\cite{BGTMats}, and others utilise GFET-based sensors as part of
a larger system, such as Hexagonfab's protein analyser~\cite{HexagonFab}. Due to their high-mobility, GFETs also have utility in analogue high frequency (HF) applications as the lack of a bandgap and resulting small on-off ratios is not an issue~\cite{Wu2011}, with Vicarelli et al. reporting GFET responsivities on the order of hundreds of GHz at room temperature~\cite{Vicarelli2012}.

As GFET research becomes more application-focused, these models 
and tools will be increasingly important. Simulation and modelling 
is a vital part in the design and development of new devices. 
Technology computer-aided-design (TCAD) is a process in the semiconductor 
industry which models the fabrication (process TCAD) and operation of 
semiconductor devices (device TCAD). The latter process models the 
behaviour of a device based on fundamental physics and comes in two 
main forms: physics-driven and compact models. The former is a more 
accurate description, though it is slow as a result, whereas the 
latter is faster, but less accurate, as the compact models often 
work one or more assumptions about device behaviour. The speed of 
compact models does however make them suitable for use with 
higher-level circuit simulation tools, such as SPICE or Verilog-A.

Our contributions in this article are as follows:

\begin{enumerate}
	\item A brief survey of existing GFET modelling solutions and 
	the motivation for our tool;
	\item Realisation of an open-source, user-friendly software tool
	for rapid simulation of GFETs to speed up the design and research 
	process;
	\item A verification of the utility and versatility of our tool 
	through comparisons to real devices with simulated devices with 
	the same geometries;
	\item A brief description of how to use the tool and how to contribute
	to further development.
\end{enumerate}

	%
%
\section{Background and Related Work}
\label{section:relatedwork}

\subsection{Background to Graphene Field-Effect Transistors}
First demonstrated by Lemme et al. in 2007~\cite{Lemme2007}, GFETs have since been the subject of an intense research focus,
as a result of their unique properties when compared to conventional metal-oxide semiconductor FETs (MOSFETs), such as their high mobilities, fast switching speeds, and ambipolarity.

The vast majority of GFETs consist of a semiconductor/oxide substrate, single- or few-layer graphene channel, produced either via chemical vapour deposition (CVD), or mechanical exfoliation, metal source and drain contacts, and either a back gate, additional dielectric and top gate, or both. Figure~\ref{figure:GFETSchems} shows these three common GFET
topologies. The first consists of a graphene channel on a semiconducting 
substrate, e.g., Si/SiO\textsubscript{2}, with source and drain contacts 
deposited on top and the doped semiconducting layer used as a back gate. 
The second is similar to the first, except with the addition of a top gate 
contact, typically separated from the graphene channel with a dielectric 
material, such as AlO\textsubscript{2}, however other dielectrics, such as ion-gels
can be used~\cite{Kim2010}.

The first topology, owing to the relatively small number of fabrication steps, is amongst the most common and many off-the-shelf GFETs feature this geometry. For sensing applications where interactions between surface adsorbents (e.g., gas molecules or proteins) and the graphene channel change the conductivity, this structure is necessary. However, graphene degrades in the air, with p-type doping as a result of moisture adsorption being a common defect. The second and third geometries, where the graphene channel is encapsulated by an insulating dielectric layer provides greater stability, although there is an impact on the electrical characteristics, namely the mobility, owing to phonon interactions between the oxides and the graphene channel. 

However, these are not the only types of GFET. Graphene nano-ribbon FETs (GNRFETs), are another form of GFET which are the subject of active research. GNRFETs are an approach to engineering a bandgap in graphene by slicing graphene into narrow ribbons. This results in a splitting of the usual 2D energy dispersion of graphene into a number of 1D modes~\cite{Chen2007}, leading to the formation of a bandgap, as some of the 1D modes do not pass through the Dirac point.

Bilayer GFETs also offer a way to induce a bandgap in graphene, with the advantage of it being electrically tunable~\cite{Zhang2009}. Such devices have a virtually identical structure to conventional monolayer GFETs, however these instead have two graphene layers. Due to the relative ease of scalable manufacture, these devices are potentially of greater interest than GNRFETs for commercialisation, however the on/off ratios are typically on the order of 100~\cite{Xia2010} and the band gaps are small, typically on the order of a few hundred meV~\cite{Zhang2009}~\cite{Xia2010}, meaning they are unlikely to be a direct replacement for silicon-based transistors in many applications.

Tunnelling GFETs form another class of device which attempts to induce a bandgap in graphene. First reported by Britnell et al.~\cite{Britnell2012}, tunnelling GFETs generally consist of a heterostructure of graphene and another GRMs such as MoS\textsubscript{2} or hexagonal boron nitride (h-BN) which acts as a tunnel barrier. Compared to e.g., bilayer GFETs, such devices can have on/off ratios as high as $10^4$, however this is still several orders of magnitude short of the standard for digital CMOS logic.

\subsection{Related Work}
There is already a substantial body of work on the simulation of GFETs, 
with Meric et al. presenting the first GFET compact model~\cite{Meric2008}
in 2008. The literature has grown to include both compact and 
physics-driven models, as well as models for different forms of GFET,
including GNRFETs. We do not intend for this article to serve as a review 
of GFET modelling, instead focusing on the use of modelling as part of the 
research and development process, so we direct the interested reader to Lu 
et al.'s comprehensive survey of GFET compact models~\cite{Lu2017}.

Currently, very few open-source, user-friendly tools for simulating 
GFETs or other devices exist. \textit{GFET Tool}, by 
Pop and Lian~\cite{PopLian2011}, is limited in the characteristics 
and properties simulated, as well as the device structure and so has 
limited use for higher-level simulations. Leong et al. developed a tool called GFETSIM~\cite{Leong2017} 
which is implemented in MATLAB and features a GUI. However, this tool 
has a different focus to that which we envisage: Leong et al.'s tool 
simulates interface charge densities in GFETs based on measured 
characteristics of real devices. Other models rely mainly on user implementation, as they provide 
SPICE-compatible models or implementations, rather than being self-contained 
tools. 
	%
%

\section{GFET Lab}
\subsection{Model Criteria}
For predicting device characteristics, it is necessary to 
identify the criteria which make a suitable model for use in GFET Lab. What follows is an overview and justification of the criteria we establish.

\subsubsection{Criteria 1: Compact Models}
Models of GFETs exist at varying levels of abstraction and Fiori and Iannaccone~\cite{Fiori2013} provide a comprehensive overview of these. We provide a brief overview below. The most complex models are \textit{ab initio} models, which typically make use of techniques such as density functional theory (DFT). These models are computationally intensive however and become very inefficient for systems larger than a few hundred atoms. The next tier of abstraction is \textit{atomistic modelling}, which predicts electronic properties of materials by defining a given system's Hamiltonian on an atomistic basis set. Increasing the level of abstraction further is \textit{semiclassical device modelling}, which is based on the Poisson equation and a transport equation (e.g., drift-diffusion model or Boltzmann transport equation), where parameters are determined either based on empirically measured device characteristics or from the results of ab initio or atomistic simulations. Analytical models form the final layer of abstraction. These, as a result of the assumptions required to make the necessary simplifications, are best suited to specific devices. As a result of their simplicity, analytical models best lend themselves to circuit simulation, rather than individual device simulations and are more likely to be compatible with other modelling tools such as SPICE or Verilog-A.


\subsubsection{Criteria 2: SPICE Compatibility}
As we intend GFET Lab to form part of a larger workflow, 
the ability to export a design to be used in higher-level 
simulations is vital. Simulation tools such as SPICE and 
Verilog-A depend on compact models and so was another 
important consideration when choosing which model to implement.

\subsubsection{Criteria 3: Minimal Fitting Parameters}
Another criteria we specify is that the number of fitting
parameters in a model is kept to a minimum. 
Often, published models introduce empirical fitting parameters 
which make them better correspond to the characteristics 
of real devices. This can be an invaluable tool when 
demonstrating a proof-of-concept, e.g., when one wishes to 
explore how an array of their experimental device that may otherwise be 
prohibitively time-consuming or expensive to fabricate would 
perform in a given application. However, for the initial design 
process of a device, this would not make sense, as one cannot 
know the characteristics of a device before they fabricate it.
The one exception we make here is for the carrier mobility ($\mu$),
as this is a fundamental part of the expression for the drain 
current, but can only be determined by electrical characterisation, e.g., using a four-point probe.

In Section~\ref{section:results}, we evaluate the simulated 
characteristics of GFET Lab against a number of real devices, 
and thus give estimates of the mobilities in the accompanying tables,
for specific device geometries, however the mobility is dependent on 
many factors, including dielectric material(s), whether the graphene 
channel is grown by CVD or mechanically exfoliated,
the presence of surface impurities and also bias voltages so these
are only rough estimates.

\subsection{Limitations of GFET Lab}
Modelling of devices is rarely perfect and GFET Lab also has a 
number of inherent limitations. The first limitation is that it makes 
no distinction between whether the device one is designing will use
CVD or mechanically exfoliated graphene. Typically, the latter
is of higher quality and has more ideal properties, as well as 
less impurities, however it is also not an easily-scalable process,
unlike CVD and so the latter is more popular for this reason.

The second limitation is that models do not account for hysteresis 
in device characteristics. Many GFET transfer characteristics exhibit 
hysteresis in the form of a shift in the Dirac point position between
forward- and reverse-gate voltage sweeps. This is contributed to by 
multiple factors, including charge trapping between the graphene channel 
and the dielectric layers~\cite{Lemme2010}, capacitive gating, which 
causes a negative shift in the position of the Dirac point, 
and charge transfer which causes a positive shift~\cite{Wang2010}
in the Dirac point with respect to gate voltage.

A third limitation is that GFET models tend to be symmetrical, i.e.,
one expects the same drain current at equal distances either side of the 
Dirac point. However, in real devices, an asymmetry is often present, 
typically for positive gate voltages, i.e., the region dominated by p-type 
conductivity. Because this asymmetry is again the result of a number of 
contributing factors, it is hard to precisely quantify or predict without 
the introduction of empirical fitting parameters. Mukherjee et al.'s model~\cite{Mukherjee2015} attempts to account for this by considering drain current as having two components, current from n-type carriers and current from p-type carriers, each with their own mobility, however this relies on one making assumptions about asymmetry.

A final limitation is the breadth of models considered. As discussed in Section~\ref{section:relatedwork}, a number of classes of GFETs exist, including GNRFETs, Bilayer GFETs, Tunnelling GFETs, those using electrolytes or ionic gels as gates, and those with functionalised graphene channels. GFET Lab does not presently include models for such devices. At the time of writing, we are aware of models for a range of devices, such as Karimi et al.'s solution-gated GFET biosensor model~\cite{Karimi2014}, Mackin's model for electrolyte-gated GFETs~\cite{Mackin2014}, Cheli's model for a bilayer GFET~\cite{Cheli2009} and Guo's review of GNRFET models~\cite{Guo2012}. However, as we discuss in Section~\ref{section:usage}, we intend for GFET Lab to be a tool contributed to by the community, making it open source so that individuals and groups can tailor it to their specific needs and share these developments with the wider research community.

\subsection{The Implemented Model}
Lu et al.'s survey of compact GFET models~\cite{Lu2017} formed a useful 
starting point to determine which model(s) to implement. Certain models 
built upon previous models, forming a chronology. For example, Thiele~\cite{Thiele2010} 
builds upon the early model of Meric~\cite{Meric2008}, however Thiele's 
model is not compact and thus not SPICE-compatible. Fregonese~\cite{Fregonese2012} 
derived a compact solution to Thiele's model, from which Rodriguez~\cite{Rodriguez2014} 
provided a further simplification to aid with hand-calculations.

The criteria outlined above helped limit our search and we further 
limited our search by not considering models for GNRFETs, bilayer GFETs, or 
tunnelling GFETs. However, there remained a large number of compact models 
which exist in the literature and so we skip providing a comparison 
of each and our resulting justification for the model's inclusion or not 
for brevity. In general, we selected models with minimal fitting parameters, 
lower computational load, and which appeared to accurately predict the 
characteristics of real devices. We also verified the dimensional consistency 
of models, to ensure that the user-defined parameters give correct output 
by performing internal conversions where necessary (e.g., from meV to Joules).

\begin{table*}[h]
	\caption{Equations for the Jimenez model~\cite{Jimenez2011}.}
	\centering
	\begin{tabular}{c|c|c}
		\toprule
		\bf{Term} & \bf{Equation} & \bf{Meaning (Units)} \\
		\midrule
		$I_D$ & $\frac{\mu_\mathrm{av} k}{2}\frac{W}{L_\mathrm{eff}}\lbrace g\left(V_{c}\right) \rbrace^{V_\mathrm{cd}}_{V_\mathrm{cs}} $ & Drain current (A)\\ 
		$L_\mathrm{eff}$ & $L + \mu_\mathrm{av} \frac{V_\mathrm{ds}}{v_\mathrm{sat}}$ & Effective channel length ($\mu$m)\\ 
		$g$ & $\left(\frac{-V_c^3}{3}\right) - \mathrm{sign}\left(V_c\right) \frac{kV_c^4}{4\left(C_t-C_b\right)} $ & Channel conductance (S) \\ 
		$V_c$ & $\lbrack \alpha \rbrack \frac{-\left(C_t+C_b\right) + \sqrt{\left(C_t+C_b\right)^2 \pm 2k \lbrack \left(V_\mathrm{gs}-V_\mathrm{gs0}-V\right) C_t + \left(V_\mathrm{bs}-V_\mathrm{bs0}-V\right) C_b } \rbrack}{\pm k}$ & Channel voltage (V) \\ 
		$k$ & $\left(\frac{2q^2}{\pi}\right) \left(\frac{q}{\left(\hbar \nu_{F}\right)}\right)$ & Constant term ($\frac{C^3}{Jm}$) \\
		\bottomrule
	\end{tabular}
	\vspace{-0.2in}
	\label{table:Jimenez}
\end{table*}

At the time of writing, we have implemented Jimenez et al.'s model~\cite{Jimenez2011} (Table~\ref{table:Jimenez}), as it appeared to best meet the criteria we outline above. We made several modifications to the model. Firstly, Jimenez et al.'s model does not account for the saturation velocity, $v_\mathrm{sat}$ and so we use the following expression from Mukherjee et al.'s model to replace the Fermi velocity, $v_F$ in the expression for the effective channel length, $L_\mathrm{eff}$, in the Jimenez model:

\begin{align}
	v_\mathrm{sat} = \frac{\omega}{\sqrt{\frac{\pi |Q_\mathrm{net}|}{q} +\frac{n_\mathrm{pud}}{2}}},
\end{align}

where $\omega$ is derived from the expression for the surface phonon energy, $E$, ($E=\omega\hbar$), $Q_\mathrm{net}$ is the net charge in the channel, given by $Q_\mathrm{net} = \beta V_c| V_c|$, and $n_\mathrm{pud}$ is a constant which accounts for charge puddles that arise due to the spatial inhomogeneity of the graphene sheet, i.e., as a result of the substrate roughness and is given by:

\begin{align}
	n_\mathrm{pud} = \frac{\Delta^2}{\pi\left( v_F \hbar \right)^2}.
\end{align}
Here, $\Delta \approxeq 54$\,meV and represents the electrostatic potential when $v_F$=1.3$\times$10\textsuperscript{6}~\cite{Zhu2009}. To account for the asymmetry typically observed in GFET transfer characteristics, we also introduce a dimensionless parameter, $\alpha=\frac{\mu_n}{\mu_p}$, which is simply the ratio of the n- to p-type mobilities ($\mu_n$ and $\mu_p$ respectively). This applies when the dominant carrier type is n-type, i.e., for $V_\mathrm{gs}-V_\mathrm{gs0}-V_\mathrm{ds} + V_\mathrm{bs}-V_\mathrm{bs0}-V_\mathrm{ds}>0$ ($V_\mathrm{gs}=$ gate-source voltage, $V_\mathrm{gs0}=$ gate-source threshold voltage, $V_\mathrm{bs}=$ back gate voltage, $V_\mathrm{gs0}=$ back gate threshold voltage, and $V_\mathrm{ds}=$ source-drain voltage). Lastly, we assume the device mobility, $\mu_\mathrm{av}$, is the average of the n- and p-type mobilities: $\mu_\mathrm{av} = \frac{\mu_n + \mu_p}{2}$.

Lee et al.'s model for a single back-gated monolayer GFET (as in Figure~\ref{figure:GFETSchems}\textbf{(a)}) appeared to be the only such model in the literature~\cite{Lee2012}, however, as it does not account for quantum capacitance, we have decided to exclude it from GFET Lab. Thus, there are presently no models for such a device structure. Instead, we provide the option for one to select air as the back-gate dielectric material ($\epsilon_r=1$) with infinite thickness and treat the top-gate as the back gate. This effectively corresponds to a zero-valued gate capacitance, which results in a division by zero when determining the $V_\mathrm{bg0}$ term and so, in this scenario, we assume $V_\mathrm{bg0}=0$.

Many of the expressions involve standard 
physical and material constants: $\epsilon_r$ and $\epsilon_0$ are the 
relative and vacuum permittivities respectively, $q$ is the elementary 
charge, $\hbar$ is the reduced Planck constant, $\eta_F$ is the Fermi 
velocity, $k_B$ is Boltzmann's constant, $T$ is the operating temperature, 
$E_F$ is the Fermi level and $E_C$ is the conduction band energy. $N_f$ is the dopant carrier density. Table~\ref{table:Dielectrics} shows the dielectrics available in GFET Lab, as well as the associated parameters. 

\begin{table}[h]
	\caption{Common dielectrics, their relative permittivities and surface-phonon energies. All except h-BN adapted from~\cite{Konar2010}, values for h-BN adapted from~\cite{Schiefle2012}.}
	\centering
	\begin{tabular}{c|c|c}
		\toprule
		\bf{Material} & \bf{Relative Permittivity ($\epsilon_r$)} & \bf{Surface-Phonon} \\
		 & & \bf{Energy (meV)} \\
		\midrule
		SiO\textsubscript{2} & 3.9 & 59.98 \\
		SiC & 9.7 & 116 \\
		Al\textsubscript{2}O\textsubscript{3} & 12.53 & 55.01 \\
		AlN & 9.14 & 83.60 \\
		HfO\textsubscript{2} & 22.0 & 19.42  \\
		ZrO\textsubscript{2} & 24.0 & 25.02 \\
		h-BN & 5.09 & 101.6 \\
		\bottomrule
	\end{tabular}
	\label{table:Dielectrics}
\end{table}

The remaining terms relate to the device geometry: $t_\mathrm{tox}$ and $t_\mathrm{box}$ 
are the top and back gate oxide thicknesses, $W$ and $L$ are the channel 
width and length and $x$ is the position along the channel. $C_{t}$ and 
$C_{b}$ are the top- and back-gate capacitances and are defined by 
$\frac{\epsilon_r \epsilon_0}{t_\mathrm{tox}}$ and $\frac{\epsilon_r \epsilon_0}{t_\mathrm{box}}$ respectively. 

Another material parameter which arises in some models is the intrinsic Fermi level of undoped graphene ($E_F$), which is reported to be around 4.57\,eV~\cite{Yu2009}, however it is also given by $E_F=\hbar\nu_{F}\sqrt{\pi n}$ for monolayer graphene~\cite{Zhu2009}, where $n$ is the carrier density. Although the model we implement does not account for these, we include them in GFET Lab to ensure ease of integration for user-defined models. 

The one fitting parameter that is essential for device modelling is the carrier mobility, $\mu$. This can be measured either using a four-point measurement, or calculated from electrical characterisation using the following expression:

\begin{equation}
	\mu = \frac{L}{V_\mathrm{ds}C_\mathrm{ox}W}\frac{dI_\mathrm{ds}}{dV_\mathrm{gs}}.
\end{equation}
However mobility is a key parameter for calculations of the device characteristics and so one must assume a value. The mobility is dependent on a number of factors, with shorter channels having lower mobilities, as well as CVD graphene generally having lower mobilities than mechanically exfoliated graphene~\cite{Venugopal2011}. Additionally, some models use the intrinsic carrier mobility, whereas others consider two regions of conductivity: that dominated by n-type carriers (i.e., electrons) and those dominated by p-type carriers (i.e., holes). In exfoliated graphene, these mobilities are theoretically symmetrical, however, due to imperfections such as doping arising from fabrication processes, there is often an asymmetry in their values and this is more pronounced for CVD graphene~\cite{Chen2014}. 

One limitation of the model is that it does not account for operating temperature. Modifications to account for this can be made by using the following expression for $n_\mathrm{pud}$ in place of the equation (2)~\cite{Zhu2009},

\begin{equation}
	n_\mathrm{pud} = \frac{2}{\pi\left( v_F \hbar \right)^2}\left( \frac{\Delta^2}{2} +\frac{\pi^2}{6}k_B^2 T^2 \right),
\end{equation}
however, during testing, we found limited impact so opted to use the simpler form. Temperature may also be incorporated by considering it in the contact resistance using expressions from Mukherjee et al.~\cite{Mukherjee2015} ($R_\mathrm{d/s0} = R_c + r_\mathrm{d/s0}\exp{\frac{q\Phi_B}{k_B T}}$), however, during development, we found these had a very limited influence on the simulated device characteristics. We also do not account for contact resistance and assume that $V_\mathrm{gs}$ and $V_\mathrm{ds}$ are intrinsic, however these are related to the extrinsic drain-source and gate-source voltages, $V_\mathrm{gs\textunderscore ext}$ and $V_\mathrm{ds\textunderscore ext}$, by:

\begin{equation}
	V_\mathrm{ds} = V_\mathrm{ds\textunderscore ext} - I_\mathrm{ds}  \left(R_d +R_s \right),
\end{equation}

and
\begin{equation}
	V_\mathrm{gs} = V_\mathrm{gs \textunderscore ext} - I_\mathrm{ds}R_s.
\end{equation}

Note that $V_\mathrm{ds}$ and $V_\mathrm{gs}$ are functions of $I_\mathrm{ds}$ and $I_\mathrm{ds}$ is a function of $V_\mathrm{ds}$ and $V_\mathrm{gs}$, meaning that these would typically need to be solve self-consistently, e.g., using the Newton-Raphson method. However, by considering the voltages used in the simulations as the intrinsic voltages, one can rearrange the above expressions for the extrinsic voltage and calculate the extrinsic voltages from the values of contact resistance and simulated drain current and plot the drain current against the calculated extrinsic voltage. In practise, this means one assumes that the $V_\mathrm{ds}$ and $V_\mathrm{gs}$ values used in the simulation are intrinsic and, once $I_\mathrm{ds}$ values are simulated, these are combined with the intrinsic $V_\mathrm{ds}$ or $V_\mathrm{gs}$ values in the above expressions to calculate the extrinsic $V_\mathrm{ds}$ and $V_\mathrm{gs}$ values.  One then plots the simulated $I_\mathrm{ds}$ values against the calculated extrinsic $V_\mathrm{ds}$ or $V_\mathrm{gs}$ values. Although Verilog-A and SPICE models have some capacity to perform Newton-Raphson iteration, this is typically used for finding operating points of circuits, rather than component simulation. Thus, the above approach makes the implementations both simpler to compute and also SPICE and Verilog-A compatible.
	%
%

\section{Simulation Evaluation}
\label{section:results}

\subsection{Model Validation}
\begin{table*}[h]
	\caption{Parameters of the experimental GFETs used for model evaluations.}
	\centering
	\begin{tabular}{l|c|c|c|c|c}
		\toprule
		\bf{Parameter} & \bf{GFET 1}~\cite{Tye2020} & \bf{GFET 2~\cite{Graphenea}} & \bf{GFET 3~\cite{Graphenea}}  & \bf{GFET 4~\cite{Graphenea}}  & \bf{Unit} \\
		\midrule
		Channel Length ($L$)                                 &   120   &  10 & 180 &  30 & $\mu$m \\
		Channel Width ($W$)                                 &   40     &  50 & 50 &  10 & $\mu$m  \\
		Top Oxide Thickness ($t_\mathrm{tox}$)	             &   50     &   90    &  90    &   90    & nm \\
		Bottom Oxide Thickness ($t_\mathrm{tox}$)        &   50     &  1  &  1  &  1  & nm \\
		Top Oxide Permittivity ($\epsilon_r$)      & 9.14  &    3.9   &    3.9   &    3.9   & -	\\
		Bottom Oxide Permittivity ($\epsilon_r$) & 9.14  & 1  & 1  & 1  & -	\\
		n-type Carrier Mobility ($\mu_n$)            & 7700   & 39 &  2400   &  4800 & cm\textsuperscript{2}V\textsuperscript{-2}s\textsuperscript{-1}	\\
		p-type Carrier Mobility ($\mu_p$)            & 8400  & 43 &  2900  & 6500 & cm\textsuperscript{2}V\textsuperscript{-2}s\textsuperscript{-1}	\\
		\bottomrule
	\end{tabular}
	\label{table:DeviceParams}
\end{table*}

\begin{figure*}
	\centering
	\subfigure[]{\includegraphics[trim=0cm 0cm 0cm 0cm, clip=true, width=0.9\linewidth]{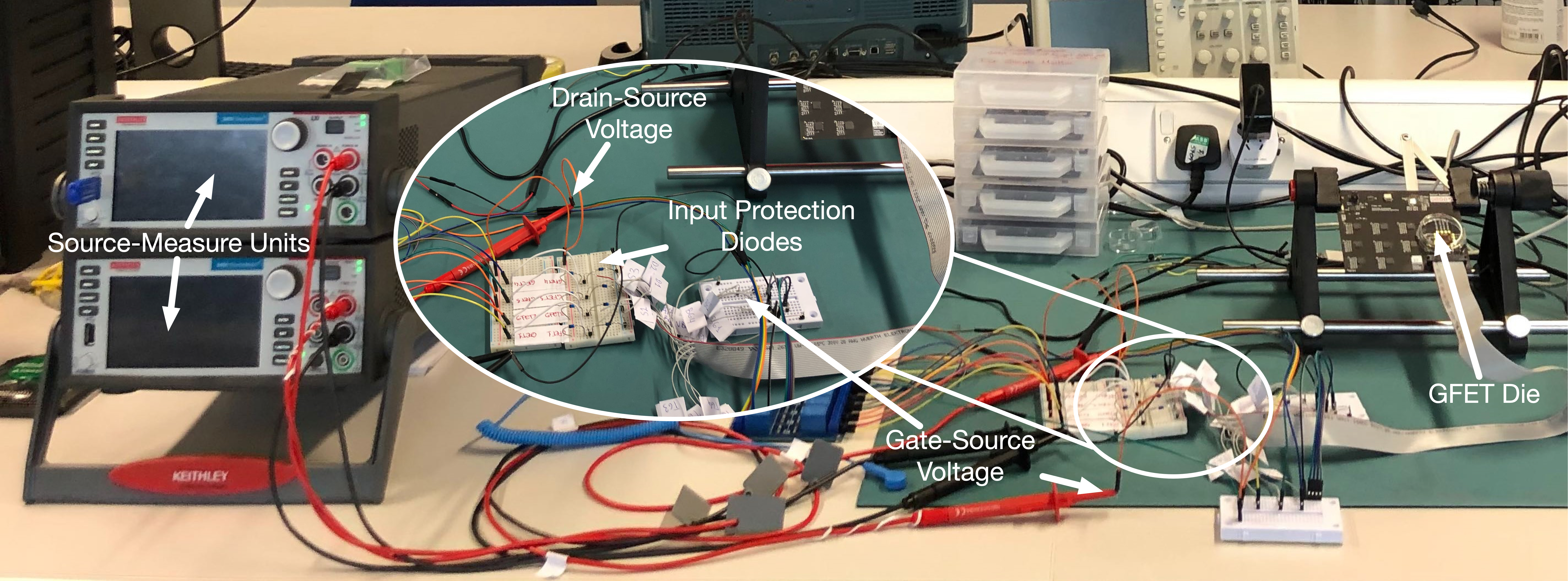}}
	\subfigure[]{\includegraphics[trim=0cm 0cm 0cm 1cm, clip=true, width=0.45\linewidth]{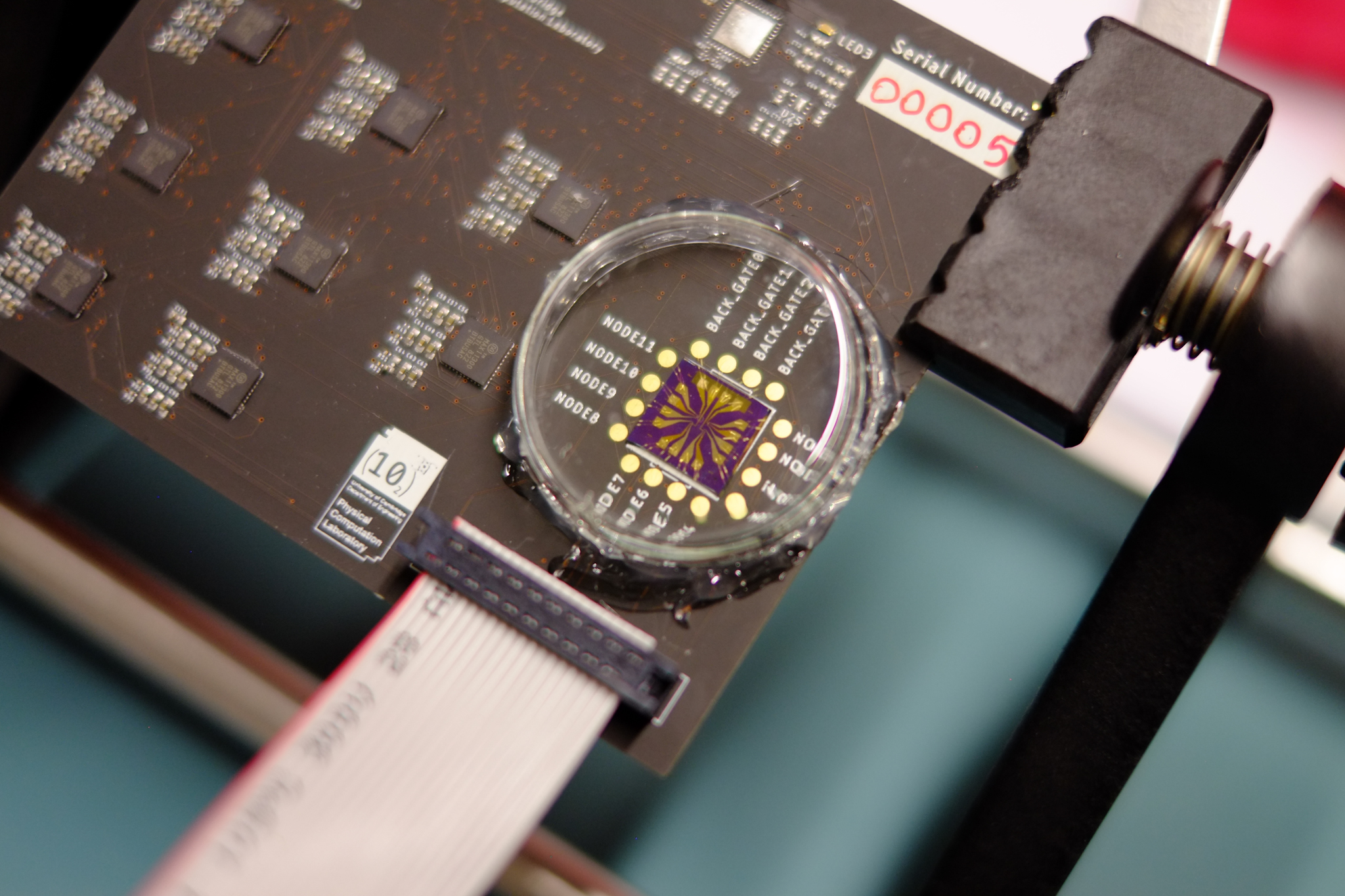}}
	\subfigure[]{\includegraphics[trim=5cm 1cm 25cm 25cm, clip=true, width=0.45\linewidth]{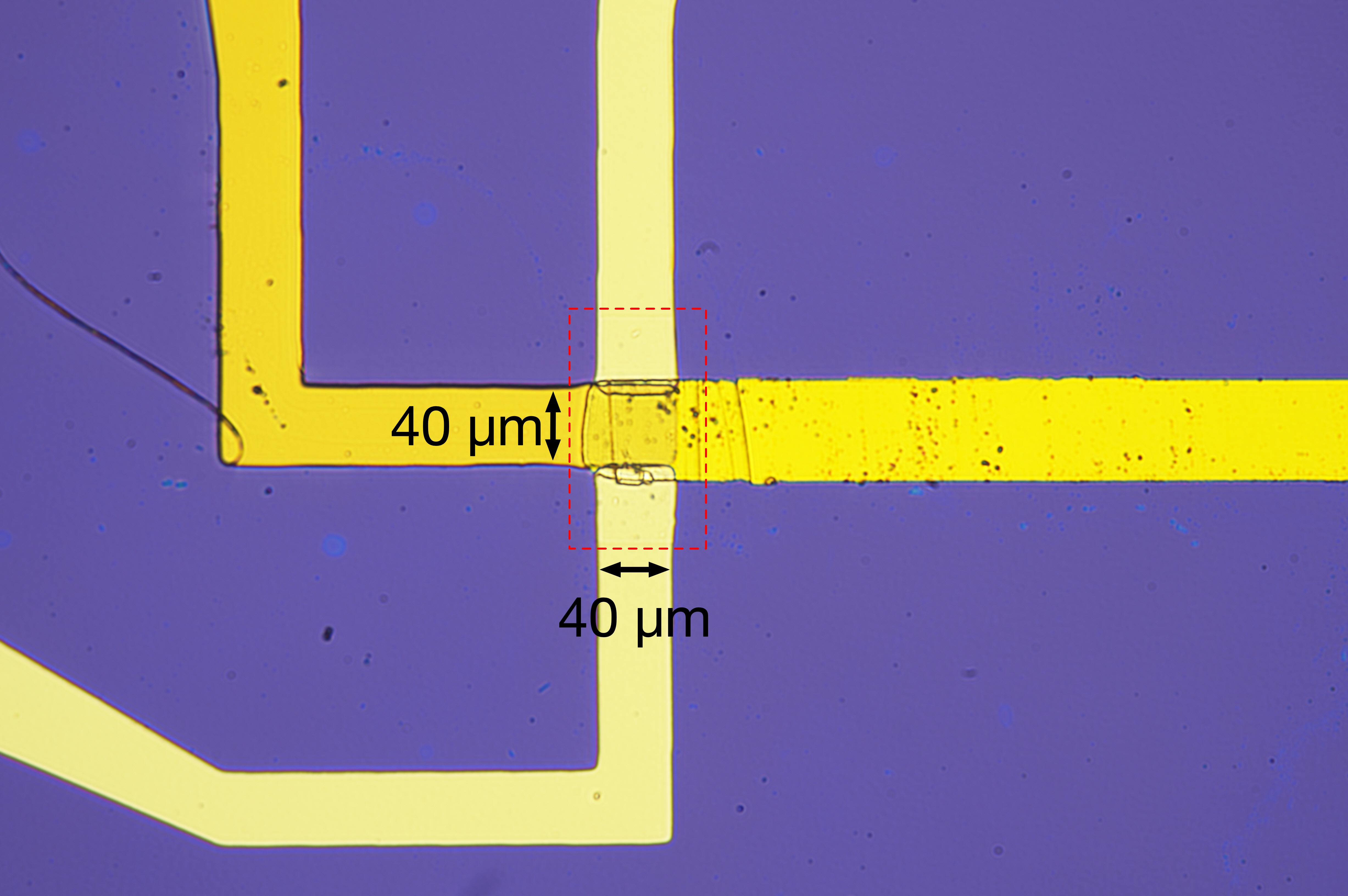}}
	\caption{\textbf{(a)} a photograph of the experimental setup used to obtain the GFET characteristics in Figure~\ref{figure:ToolVsData}\textbf{a} and \textbf{b}. The inset shows the input-protection circuit on a breadboard consisting of two Zener diodes, via which the GFET is connected to the SMUs; \textbf{(b)} a photograph of the GFET die in Figure~\ref{figure:GFETSchems} \textbf{(d)} wire-bonded to the experimental platform; \textbf{(c)} microscope image of the GFET whose characteristics we present in Figure~\ref{figure:ToolVsData}\textbf{(a)} and \textbf{b}. We presented \textbf{(b)}, \textbf{(c)}, and \textbf{(d)} in a previous article~\cite{Tye2020}.}
	\label{figure:ExperimentalSetup}
\end{figure*}

\begin{figure*}[h]
	\centering
	\subfigure[]{\includegraphics[trim=4cm 7cm 5cm 8cm, clip=true, width=0.4\linewidth]{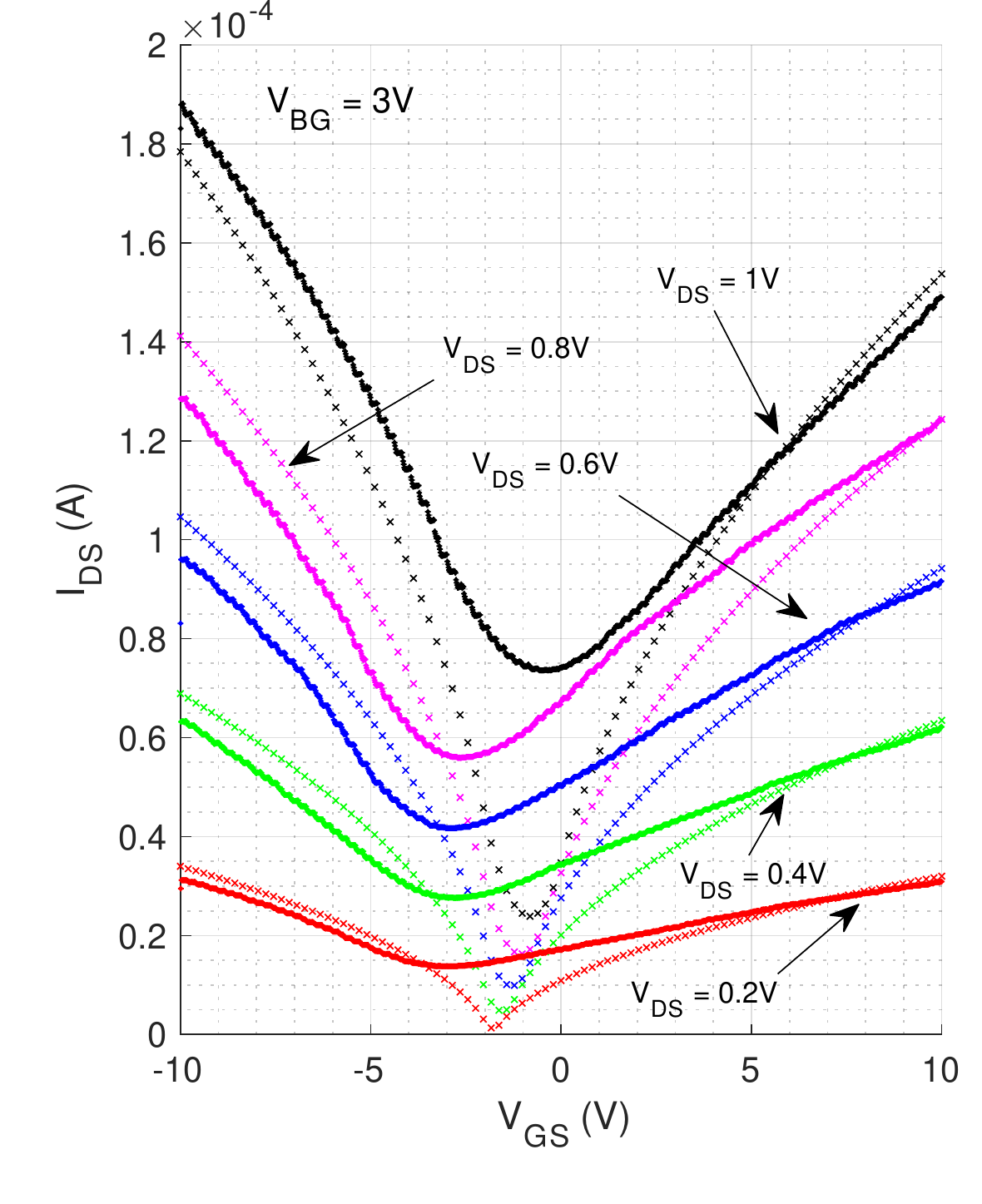}}
	\subfigure[]{\includegraphics[trim=4cm 7cm 5cm 8cm, clip=true, width=0.4\linewidth]{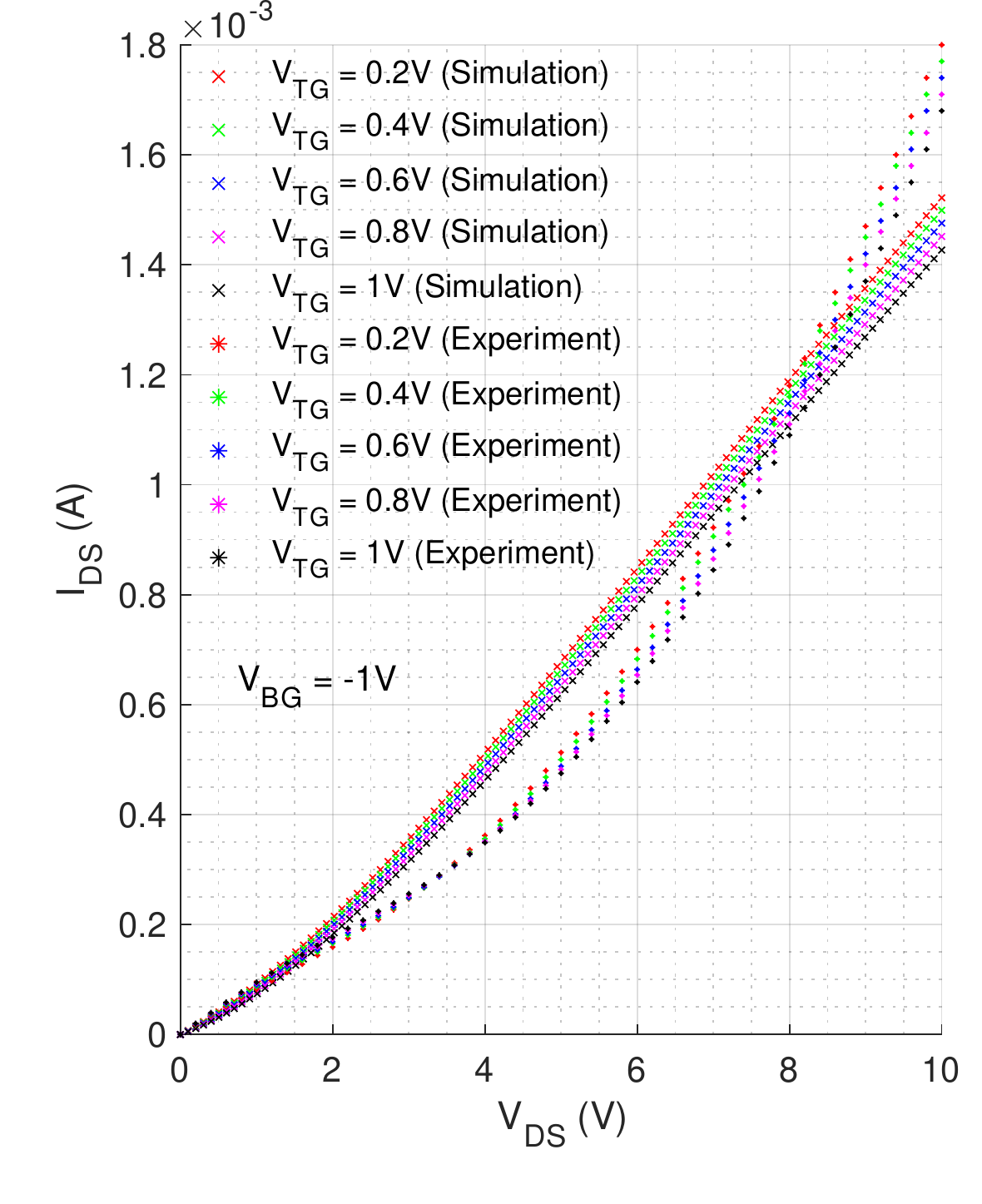}}
	\subfigure[]{\includegraphics[trim=3cm 7cm 4cm 8cm, clip=true, width=0.3\linewidth]{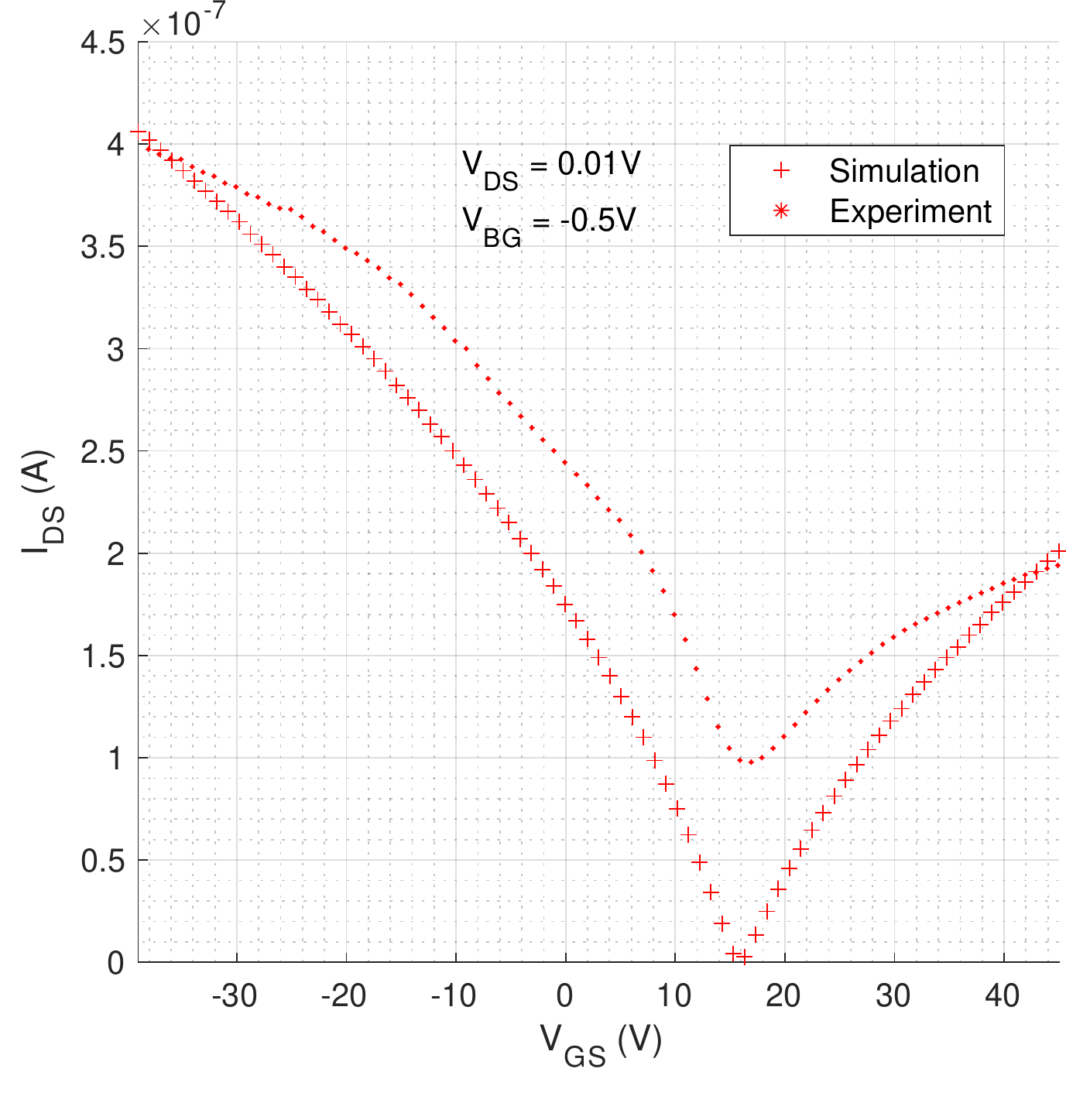}}
	\subfigure[]{\includegraphics[trim=3cm 7cm 4cm 8cm, clip=true, width=0.3\linewidth]{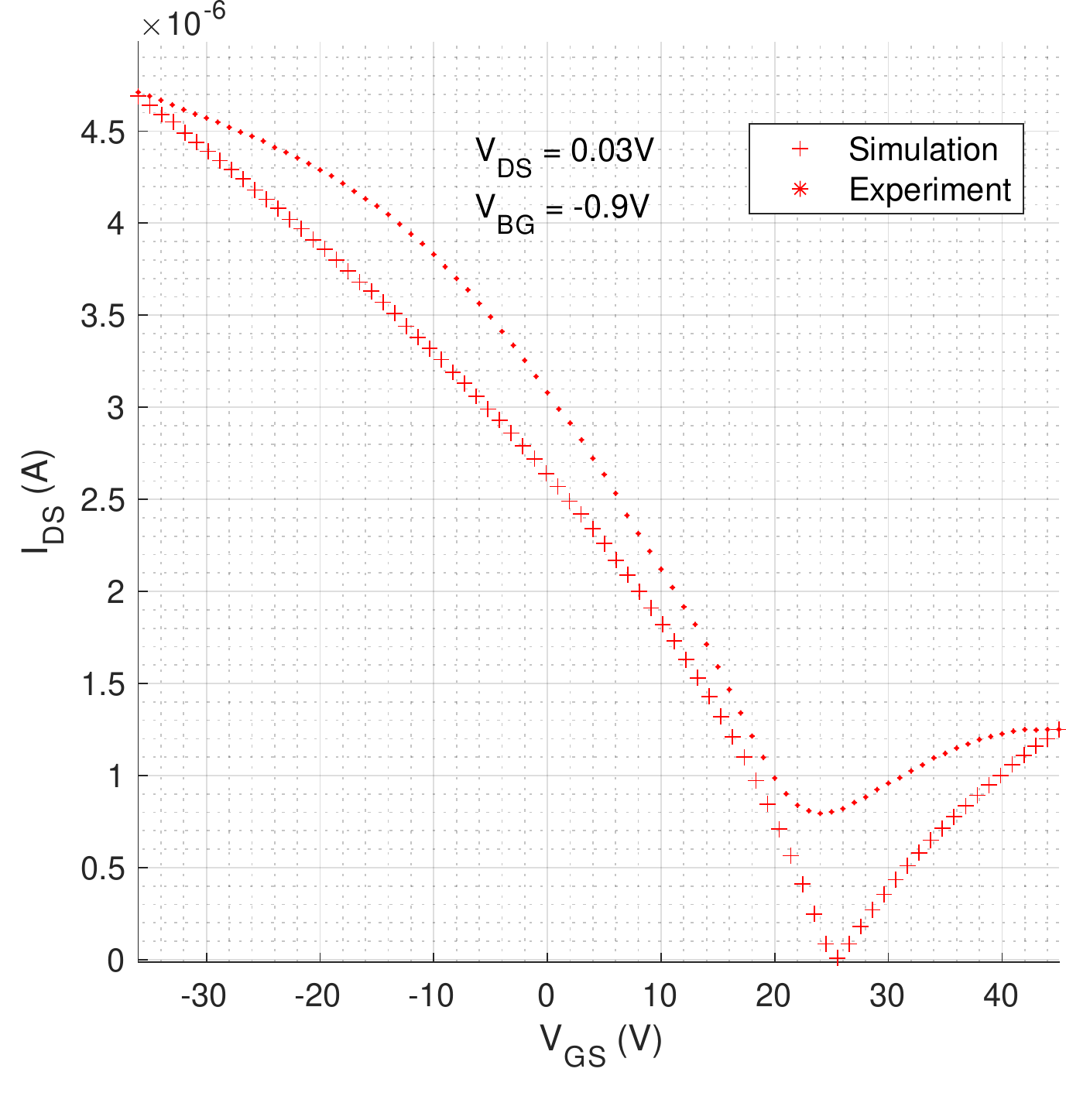}}
	\subfigure[]{\includegraphics[trim=3cm 7cm 4cm 8cm, clip=true, width=0.3\linewidth]{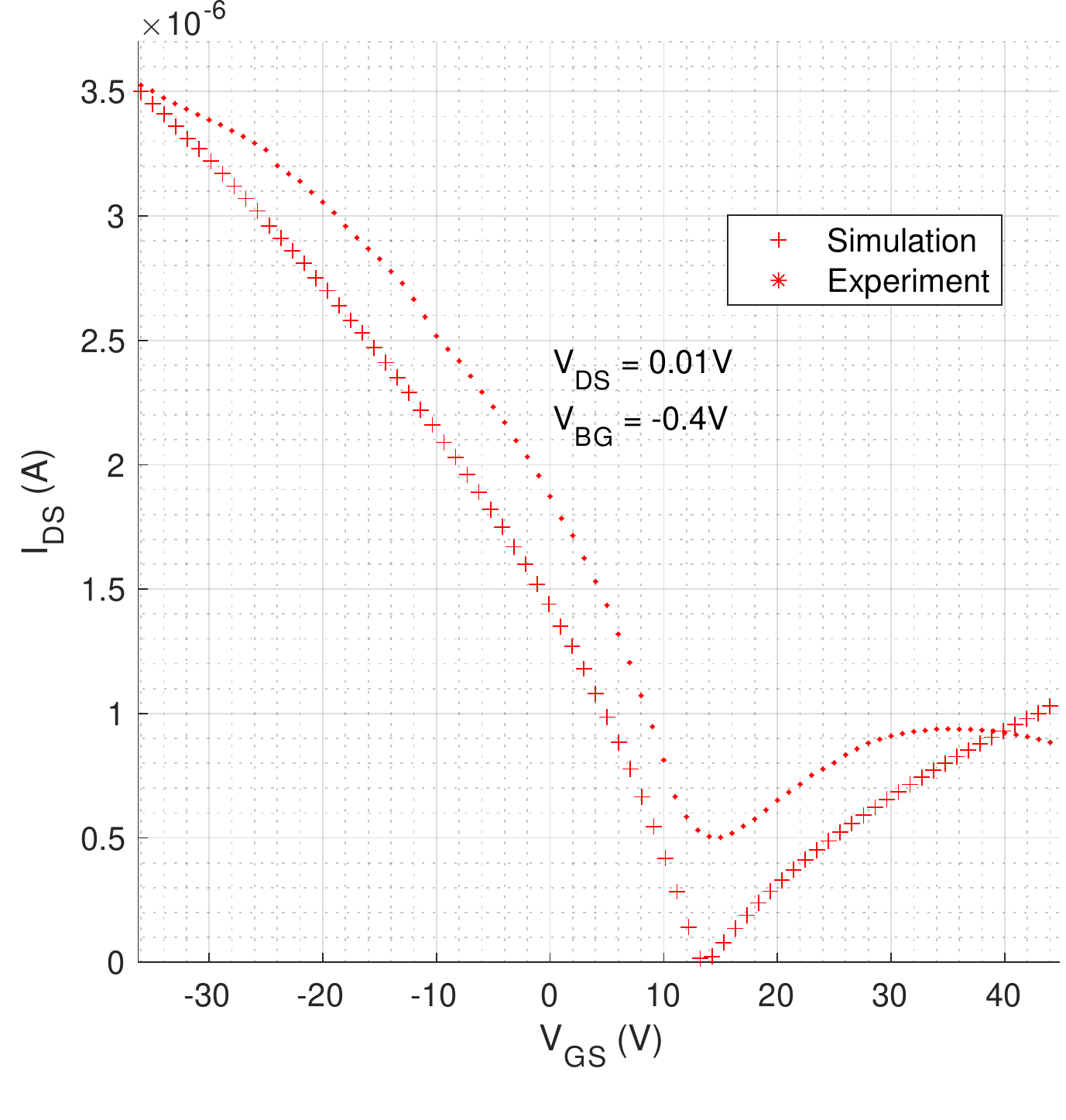}}
	\caption{Comparison of simulated and experimentally measured GFET transfer (\textbf{(a)}) and I-V (\textbf{(b)}) characteristics. \textbf{(c)}-\textbf{(e)}: Comparison of simulated and experimentally measured GFET transfer characteristics for several GFETs from the Graphenea S10. In all cases, the stars represent experimental data and the crosses represent the simulation results.}
	\vspace{-0.1in}
	\label{figure:ToolVsData}
\end{figure*}

To verify the applicability of our tool, it is worthwhile to 
compare the performance to real GFETs. We compare a number of 
devices with different geometries, shown in Table~\ref{table:DeviceParams}, to validate the versatility of GFET Lab. Figure~\ref{figure:ToolVsData} shows comparisons of the simulated and experimentally-obtained transfer and I-V characteristics of GFETs with different channel geometries and dielectric materials. The first GFET (GFET 1) consists of an SiO\textsubscript{2} substrate with an Au back gate, a 50\,nm AlO\textsubscript{3} dielectric layer, a monolayer graphene channel with Au source/drain contacts, another 50\,nm AlO\textsubscript{3} dielectric layer and an Au top gate. We describe the fabrication and characterisation of the GFET in a previous work~\cite{Tye2020}. We then validate the tool's performance for both alternative device structures, as well as for the same fabrication process with different channel dimensions, by comparing simulated and experimental characteristics for several devices (GFETs 2-4 in Table~\ref{table:DeviceParams}) from the Graphenea S10, an off-the-shelf die consisting of 36 graphene devices with various geometries. In all simulations, we take $N_f$ to be $5 \times 10^{15}$. We attempted a na{\"i}ve derivation of carrier mobility values from the experimental data using equation (3), however these values did not appear accurate and so, for the purposes of validation, we estimate values of mobility which provide a good fit to the data. We measured the output characteristics by measuring the GFET drain current ($I_\mathrm{ds}$) as a function of the drain-source voltage ($V_\mathrm{ds}$). Different factors in GFETs can cause poor current saturation including the lack of bandgap, interfacial phonon scattering and inadequate electrostatic control~\cite{Meric2008}~\cite{Mitta_2020}. This leads to a large region of linear output characteristics. This region is valuable for extracting various parameters, including the quality of the contact resistance. We obtained the transfer characteristics of GFET by measuring the drain current as a function of the gate voltage ($V_\mathrm{gs}$) for constant $V_\mathrm{ds}$. Resistivity (conductivity) and mobility which are important parameters of GFETs can be extracted from transfer characteristics. We performed electrical characterisation of the Graphenea S10 devices using a Keithley 4200 semiconductor characteristic analyser system (Keithley, 4200 SCS) combined with a probe station in the air.

Figures~\ref{figure:ToolVsData} \textbf{(a)} and \textbf{(b)} show the experimentally-obtained and simulated transfer and I-V characteristics for GFET 1. \textbf{(a)} shows multiple transfer characteristics for the same device, with $V_\mathrm{ds}$ bias voltages ranging from 0.2 to 1\,V. The model matches the experimental data to a reasonable degree, with the orders of magnitude being relatively close to the measured data, although the on/off ratio shows some discrepancy: the on/off ratio for the experimental data ranges from 2.27 to 2.41, whereas the on/off ratios for the simulations range between 7.5 to 25. In real devices, a voltage sweep typically causes a shift in the position of the Dirac point as a result of changing charge trap densities, as accounted for by $n_\mathrm{pud}$, influencing the mobilities and saturation velocity, however our model does not account for this and assumes a constant $n_\mathrm{pud}$. This explains in part the discrepancy in on/off ratios, as, by equation (1), $v_\mathrm{sat}$ depends on $n_\mathrm{pud}$. When the net channel charge, $Q_\mathrm{net}$ is small, i.e., when $V_c\approxeq0$,  $n_\mathrm{pud}$ becomes the dominant factor in the value of $v_\mathrm{sat}$ and so discrepancies between this value and the actual value will translate to variations in the on/off ratio.

Figure~\ref{figure:ToolVsData} \textbf{(b)} shows the IV characteristics for the same device, with $V_\mathrm{gs}$ bias voltages ranging from 0.2 to 1\,V. The simulation and experimental characteristics, although on the same order of magnitude, show a larger discrepancy. In the IV characteristics simulation, we used a $V_\mathrm{bg}$ bias of -1\,V, which gives reasonable agreement with the experiment, being of the same order of magnitude and similar values for the same bias voltages. Both characteristics show a point of inflection as expected for a GFET, however the point occurs at around 2\,V in the simulation, compared to around 4\,V in the experiment. In a real device, bias conditions will be different depending on the nature of the voltage sweep and thus influence other parameters which the model assumes are constant, such as the mobility and carrier velocity. 

Figures~\ref{figure:ToolVsData} \textbf{(c)}-\textbf{(e)} shows simulated and measured transfer characteristics for three GFETs from the Graphenea S10~\cite{Graphenea} whose details are described in Table~\ref{table:DeviceParams}. As the S10 devices have a single back gate, we treat the back gate in the model as a top gate with a 90\,nm SiO\textsubscript{2} dielectric. We account for the large shift in the Dirac point position in the experimental characteristics by adding a back gate bias voltage, $V_\mathrm{bg}$. We consider the virtual gate oxide to be 1\,nm thick with a dielectric constant equal to the vacuum permittivity. Again, the shape and order of magnitude of the simulated device is relatively close to the experimental data, however there is a discrepancy between the on/off ratios of the experimental and simulated GFETs. The experimental data have on/off ratios ranging between about 4 and 7, whereas the on/off ratios for the simulations range from about 153 to 530. 

In each device, there are discrepancies between the simulated and real device characteristics, which are likely due to non-idealities which arise in devices. For example, asymmetries in the measured characteristics versus the modelled characteristics can arise due to unintentional doping as a result of the fabrication process or device degradation, something which many models don't account for, however our introduction of the parameter $\alpha$ appears to account well for this. Additional fitting parameters and fine-tuning of the model would give a closer response, however, as our intention here is a generalised tool for predicting device characteristics, rather than fitting a model closely to a single device, these results suggest broadly good agreement with experiments. As fitting parameters are typically empirical in nature, this only really makes them useful after fabrication, e.g., when one wishes to simulate a circuit consisting entirely of fabricated devices, rather than to predict the characteristics of a device before fabrication. 

A systematic way to quantify the influence of the fabrication process would be very useful, e.g., by considering fitting parameters as \textit{process parameters}. For example, coefficients to denote CVD vs exfoliated graphene, doping as a result of the transfer and patterning processes, the influence of the growth substrate etc., with values determined by fabrication (e.g., establishing a relationship between resist etching time and doping levels in electrical characteristics), could provide a more rigorous framework for device design and simulation. 

\subsection{Integration with Higher-Level Tools}
\begin{figure*}[h]
	\centering
	\subfigure[]{\includegraphics[trim=4cm 7cm 5cm 8cm, clip=true, width=0.35\linewidth]{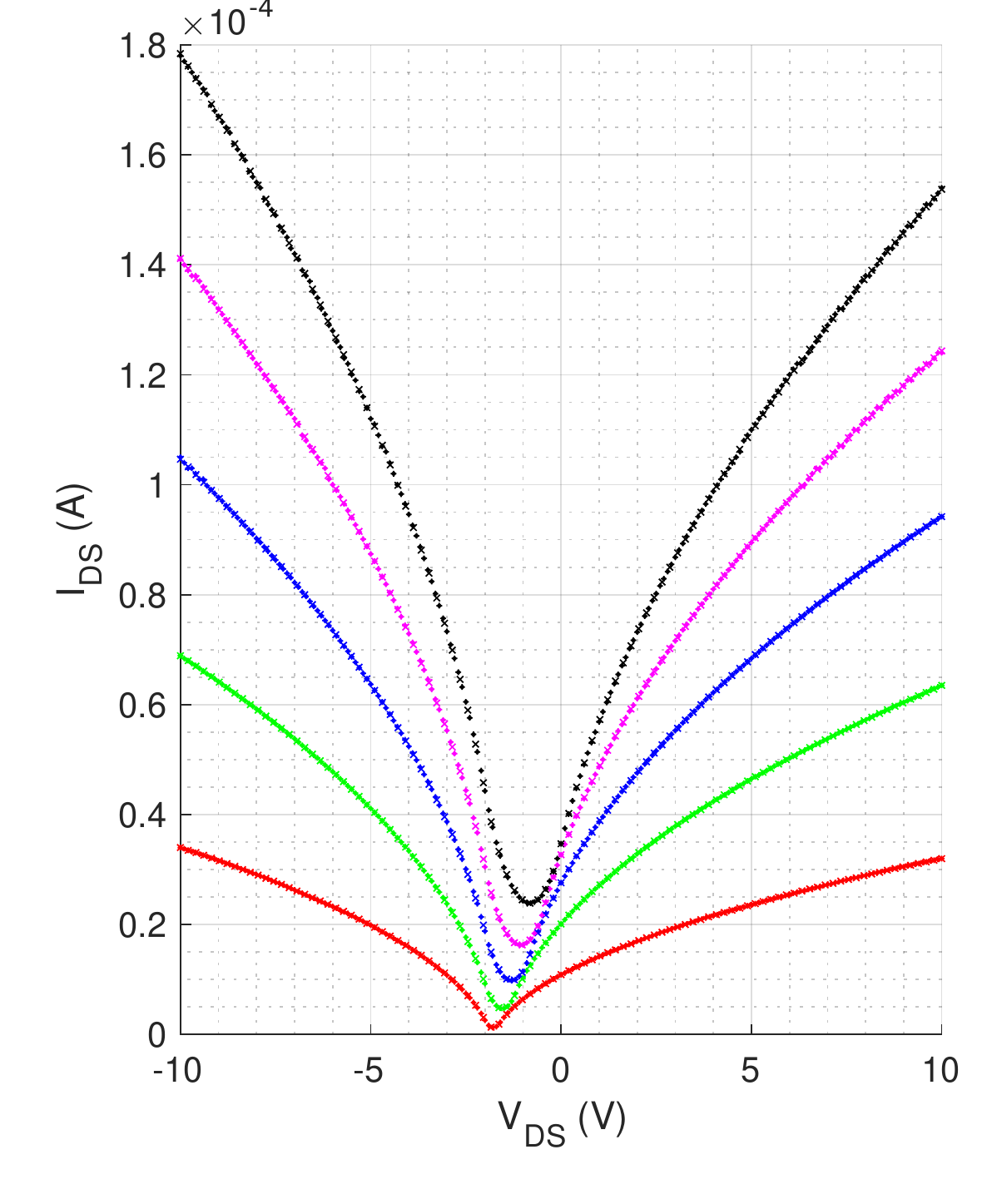}}
	\subfigure[]{\includegraphics[trim=4cm 7cm 5cm 8cm, clip=true, width=0.35\linewidth]{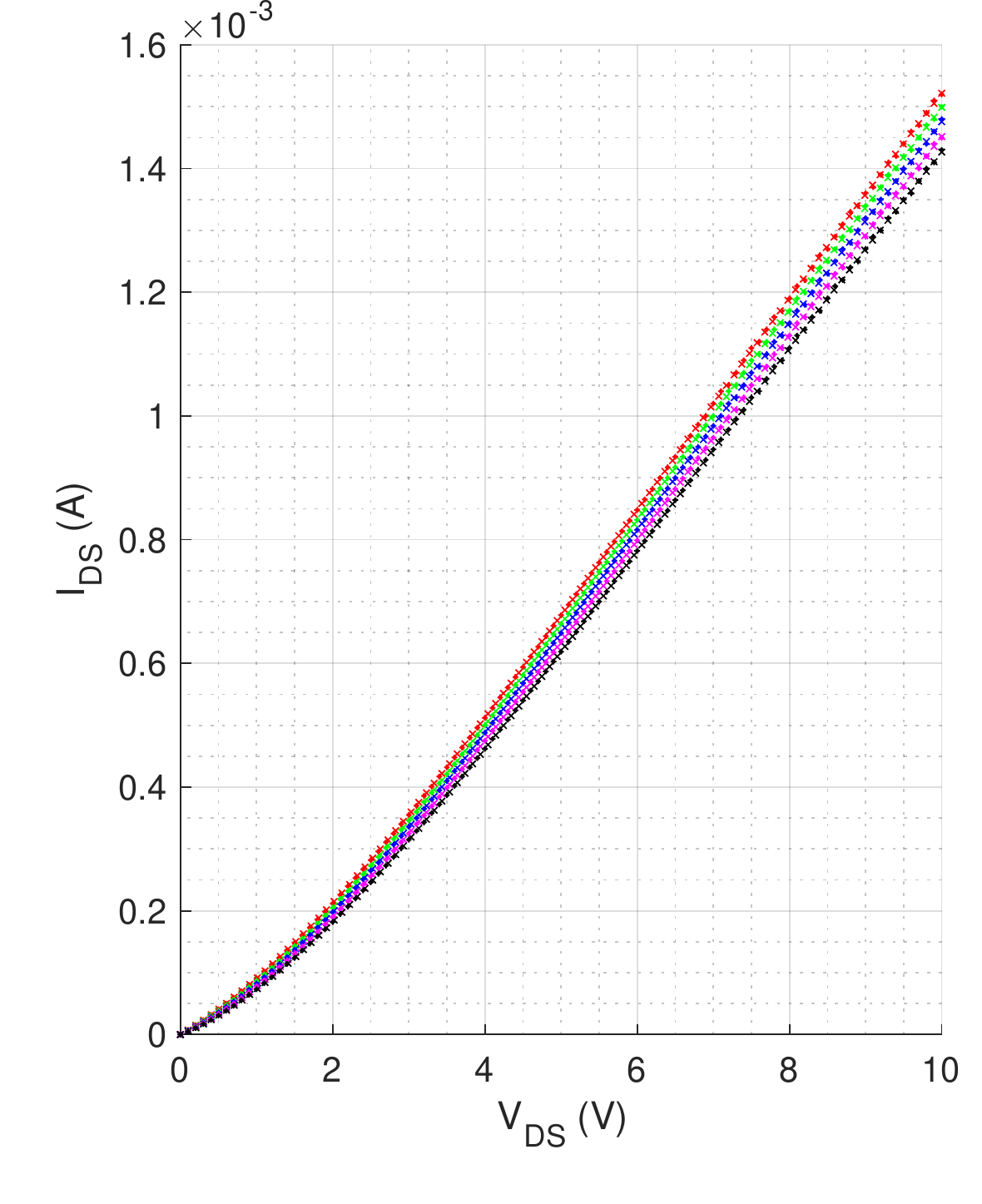}}
	\subfigure[]{\includegraphics[trim=3cm 7cm 4cm 8cm, clip=true, width=0.3\linewidth]{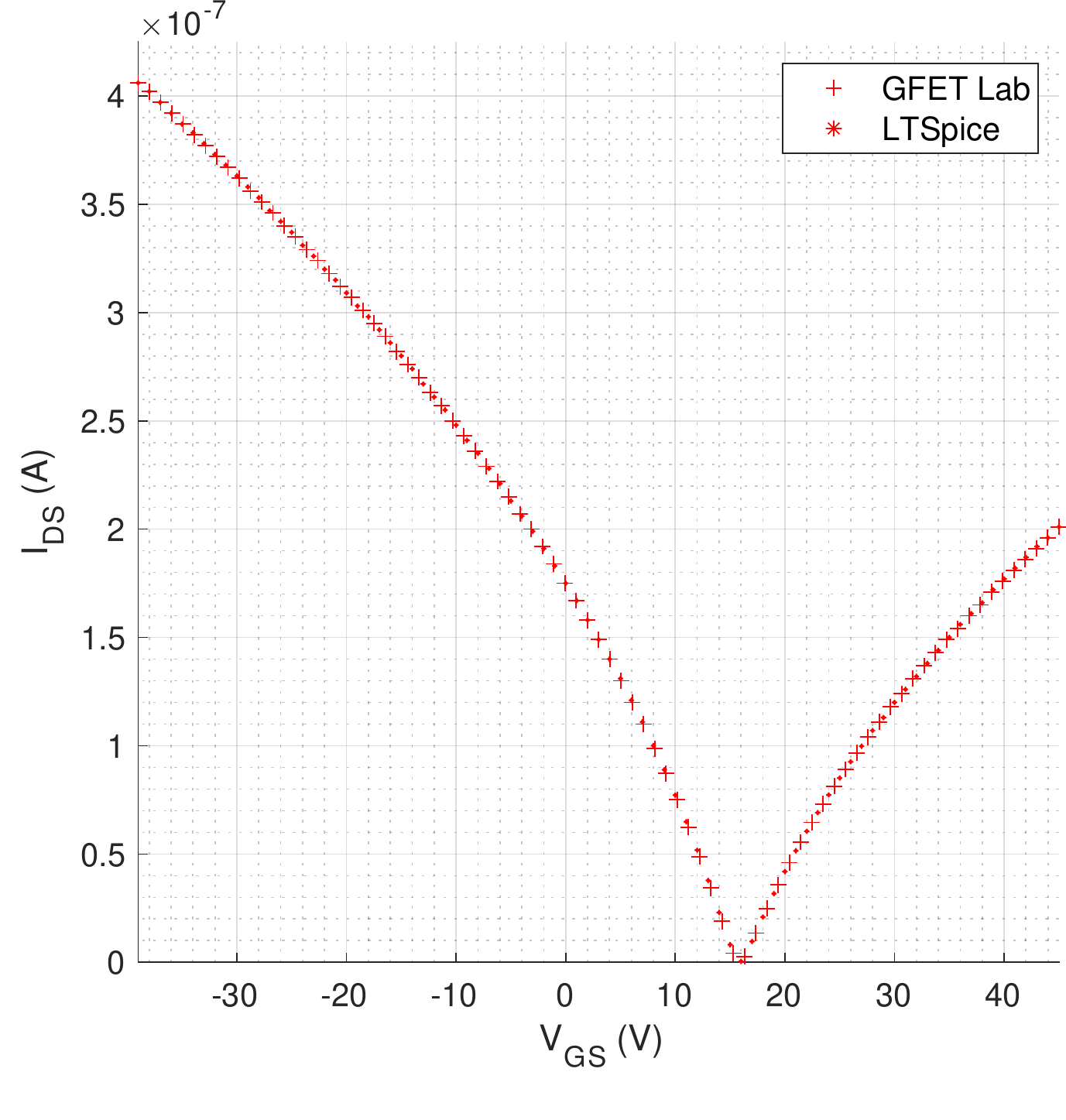}}
	\subfigure[]{\includegraphics[trim=3cm 7cm 4cm 8cm, clip=true, width=0.3\linewidth]{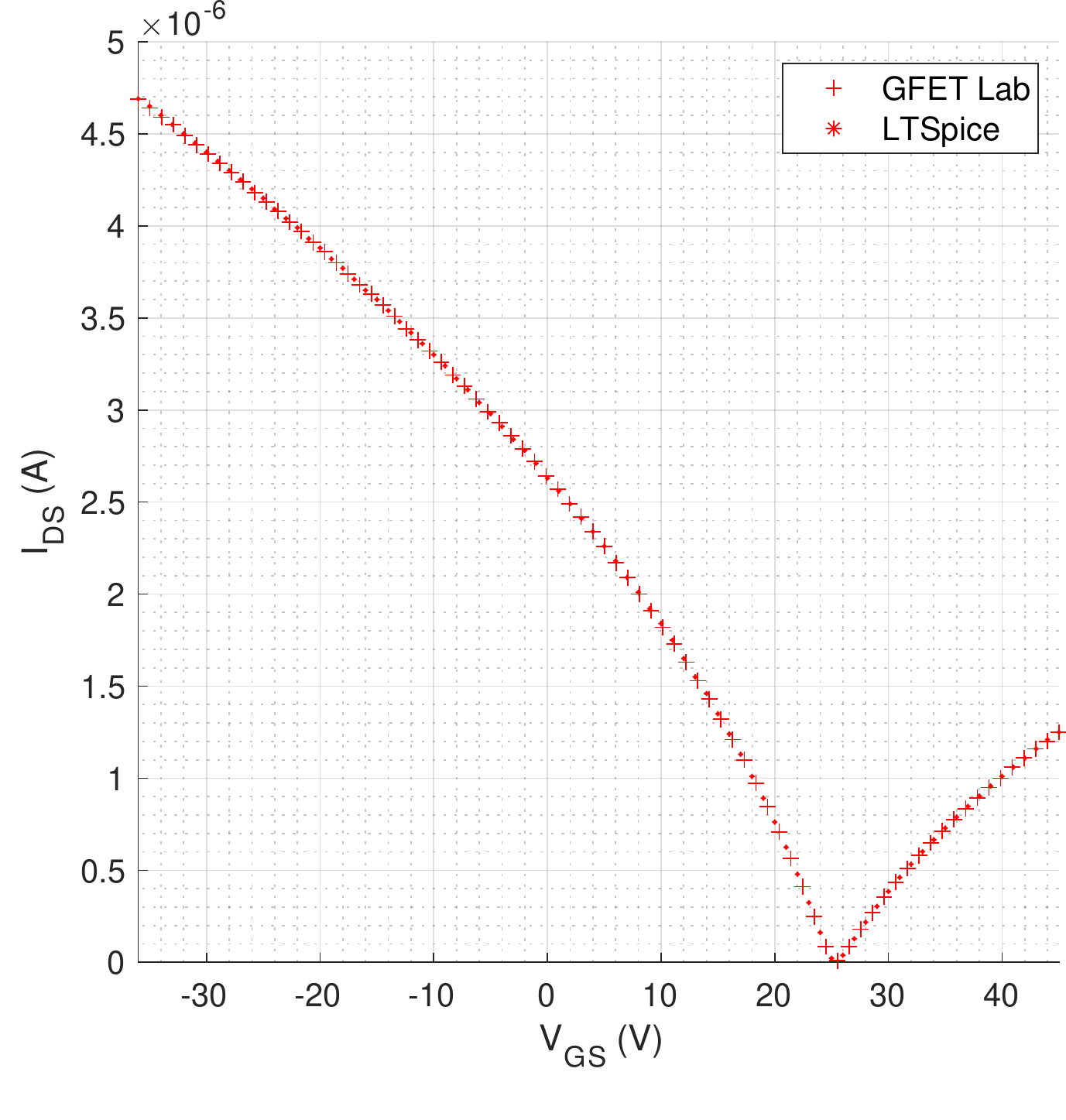}}
	\subfigure[]{\includegraphics[trim=3cm 7cm 4cm 8cm, clip=true, width=0.3\linewidth]{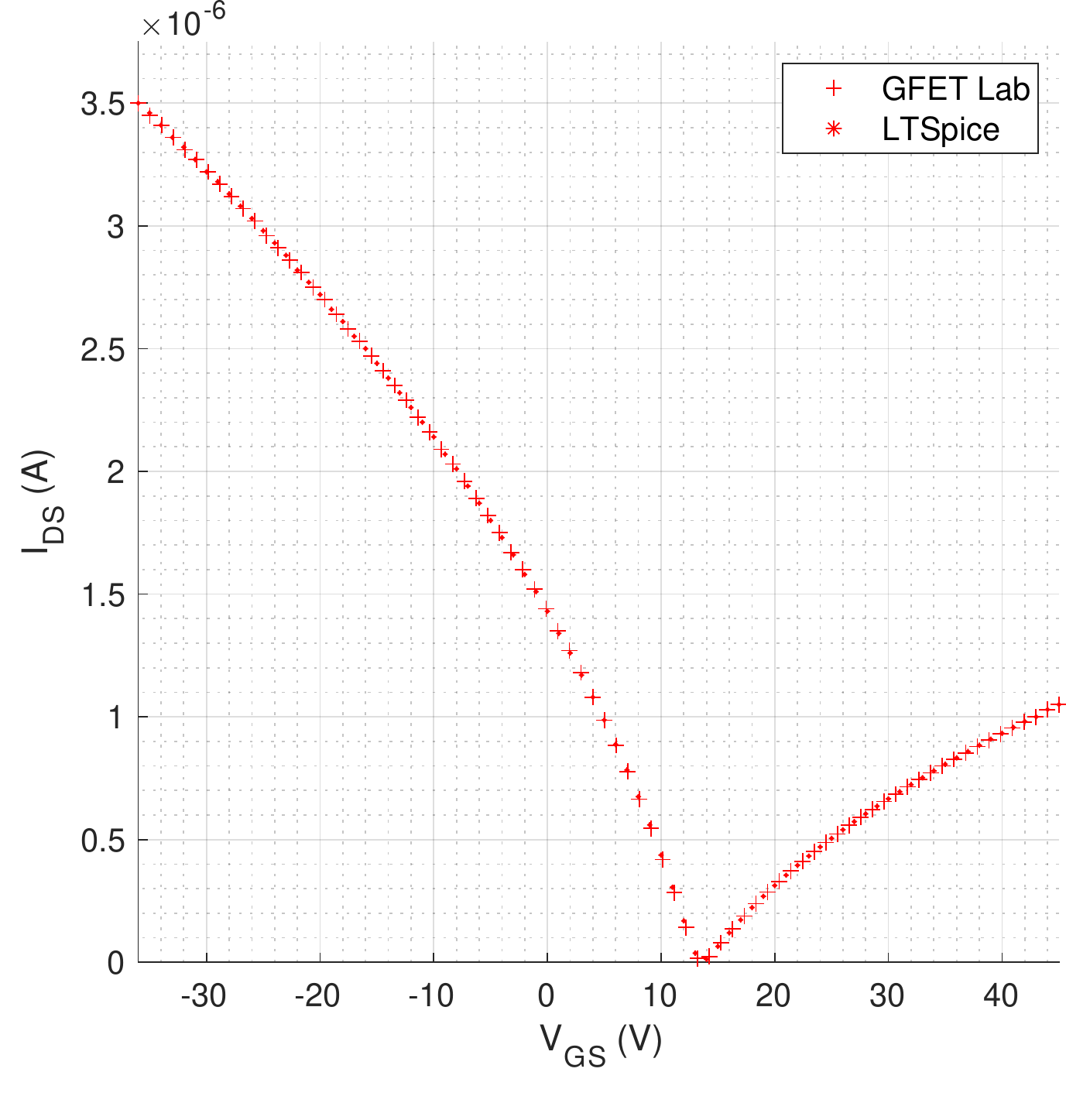}}
	\caption{
		\textbf{(a)} Comparison of the simulated transfer characteristics of GFET 1 from both GFET Lab and the exported SPICE model for $V_\mathrm{ds}=0.2-1$\,V;
		\textbf{(b)}  Comparison of the simulated I-V characteristics of GFET 1 from both GFET Lab and the exported SPICE model for $V_\mathrm{gs}=0.2-1$\,V;
		\textbf{(c)} Comparison of the simulated transfer characteristics of GFET 2 from both GFET Lab and the exported SPICE model;
		\textbf{(d)}  Comparison of the simulated I-V characteristics of GFET 3 from both GFET Lab and the exported SPICE model;
		\textbf{(e)}  Comparison of the simulated I-V characteristics of GFET 4 from both GFET Lab and the exported SPICE model. In each, the LTSpice and GFET Lab model parameters are the same as in Figure~\ref{figure:ToolVsData}. In both, crosses correspond to GFET Lab simulations and dashes correspond to LTSpice simulations.}
	\vspace{-0.1in}
	\label{figure:ToolVsSpice}
\end{figure*}

We also verify the correct output for exported SPICE models. Figure~\ref{figure:ToolVsSpice} shows the simulated transfer and I-V characteristics both in GFET Lab using the exported models in LTSpice for GFETs 1-4. As the figure shows, the LTSpice simulation and GFET Lab transfer and I-V characteristics are consistent.

	\section{GFET Lab: Usage and Open-Sourcing}
\label{section:usage}
\subsection{GFET Lab Tutorial Overview}
\begin{figure}[h]
	\centering
	\subfigure[]{\includegraphics[width=0.95\linewidth]{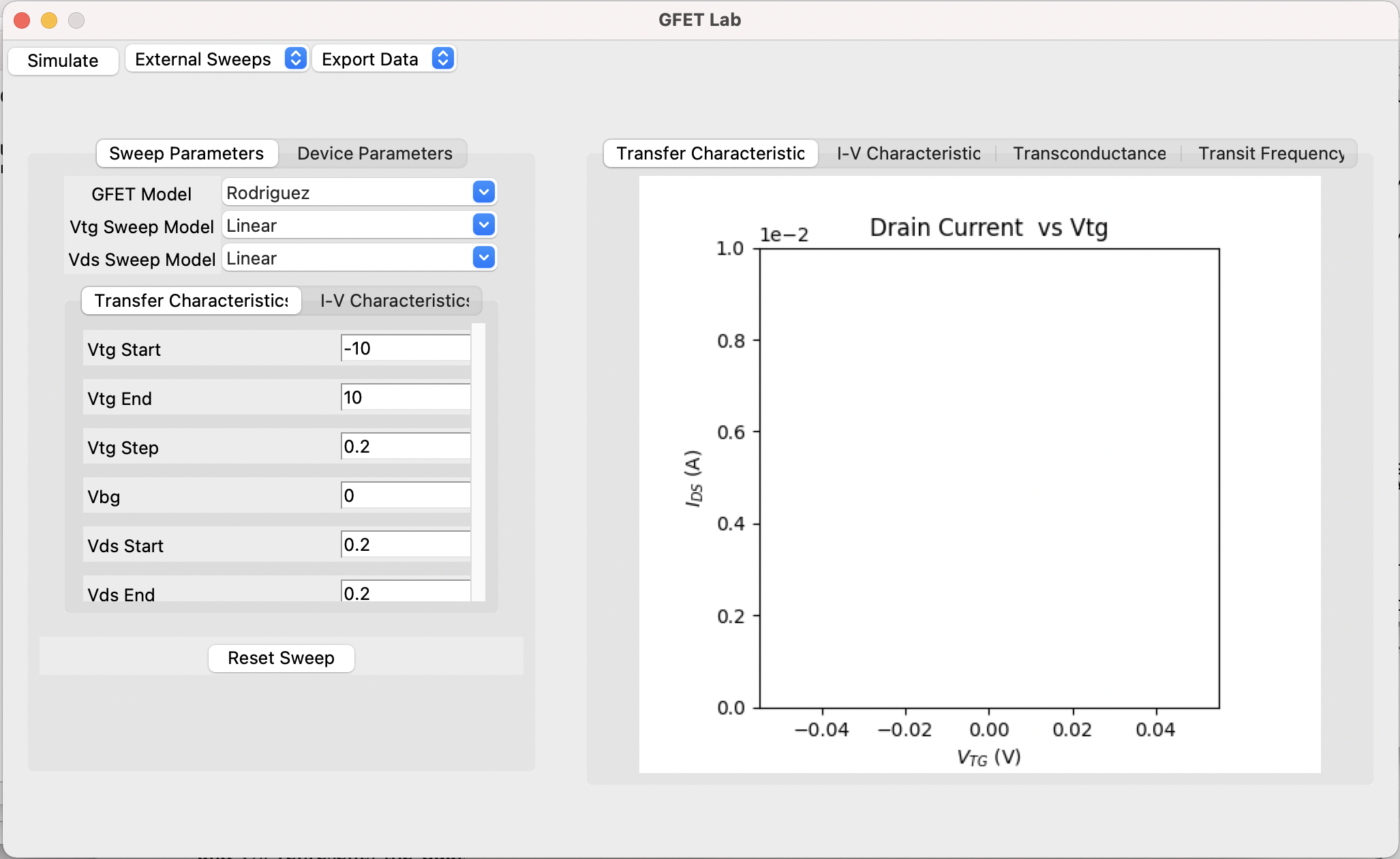}}
	\subfigure[]{\includegraphics[width=0.95\linewidth]{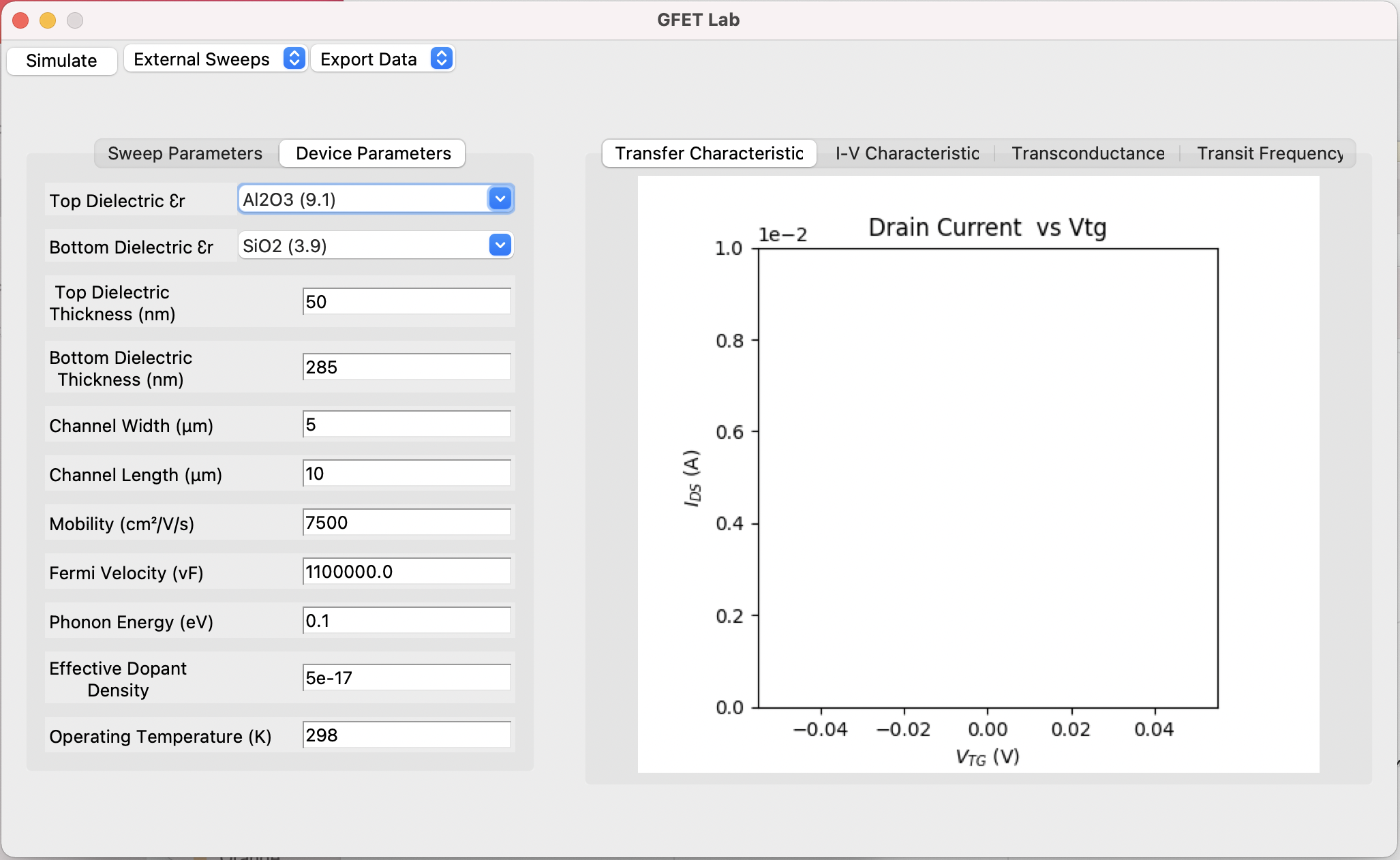}}
	\caption{\textbf{(a)} screenshot showing the main screen when the program is launched. 
		Here, the user can specify the parameters of the transfer characteristics sweep, 
		as well as selecting the model they wish to use; \textbf{(b)} screenshot 
		showing the tab where the user can specify the device parameters, such as 
		gate dielectric, channel dimensions, carrier mobility and others. The parameters 
		used will depend on the selected model and so some parameters will be ignored 
		for some models.}
	\label{figure:ToolScreenshot1}
\end{figure}

As a stand-alone software tool, GFET Lab is intended to be used exclusively through the GUI. Figure~\ref{figure:ToolScreenshot1} shows the main screen of the program. 

\begin{figure}[h]
	\centering
	\subfigure[]{\includegraphics[width=0.52\linewidth]{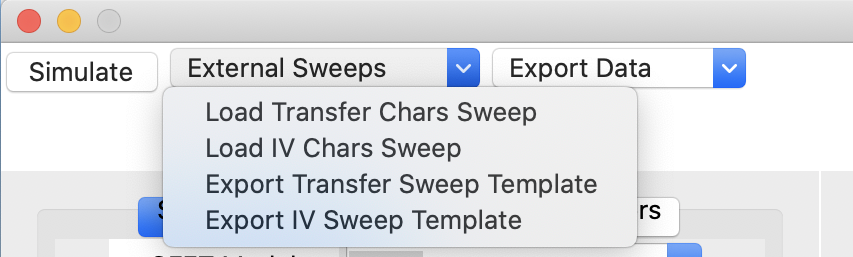}}
	\subfigure[]{\includegraphics[width=0.52\linewidth]{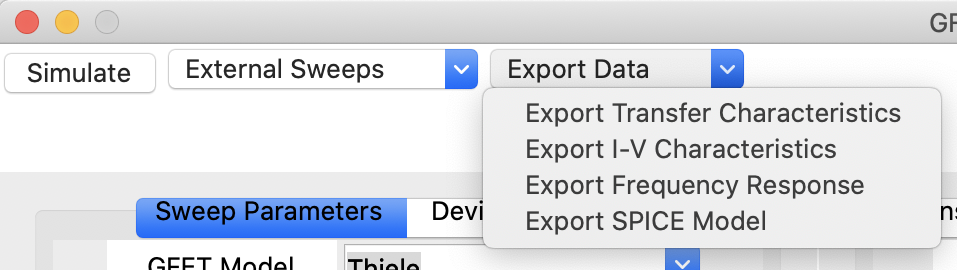}}
	\caption{These screenshots show to possible import and export options for the simulator. 
		\textbf{(a)} shows the importing options for loading external sweeps, as well as 
		allowing users to export a template sweep file for ease of use. \textbf{(b)} 
		shows the various data exporting options, including the transfer characteristics, 
		the I-V characteristics, frequency response and a SPICE model of the simulated device.}
	\label{figure:ToolIO}
	\vspace{-0.2in}
\end{figure}

At the top, a button to run the simulations, alongside drop-down menus for file I/O functionality is provided. The first menu allows users to load custom voltage sweeps, as well as exporting a template file which is in a readable format for GFET Lab. The second drop-down menu allows users to export simulation results and SPICE models. Figure~\ref{figure:ToolIO} shows the options in each menu.

The left-hand frame forms the virtual laboratory aspect of the software, where the users can specify device and sweep parameters, as well as selecting the device model they wish to use. The ``Device Parameters'' tab is where the user specifies the device geometry (dielectric thickness(es), channel dimensions), the dielectric material (e.g., AlO\textsubscript{2}, HfO\textsubscript{2}...), the operating temperature and other material parameters, such as the carrier mobility, the Fermi velocity, the surface phonon energy and the effective dopant density.  

The ``Sweep Parameters'' tab allows the user to define the GFET model to be used, as well as the voltage sweeps they wish to simulate. For example, whether to sweep the gate voltage or the source-drain voltage and the shape of the sweep (linear, logarithmic etc.). These settings are based on the functionality of the Keithley 2400 source-measure unit, a common tool for device electrical characterisation. There are two main plots used for the characterisation of transistors, a transfer characteristic plot, where the source-drain voltage is kept constant and the gate voltage is swept, and an I-V characteristics plot, where the gate voltage remains constant and the source-drain voltage is swept. As both use slightly different loops in the code, we provide both in separate tabs and a single simulation run will conduct both.

The right-hand frame displays the simulation results. The user simply selects a given tab to display the plot they wish to see. In addition to the transfer and I-V characteristics, the software also plots the transconductance and transit frequency of the simulated device, both of which are often of interest to researchers. 

\subsection{Community Contribution}
We developed the tool with two main motivations. 
Firstly to help in the development of GFET circuits, by allowing for rapid simulation 
of GFET characteristics with different parameters. These include bias voltages and 
device structures (i.e., gate oxide thickness, oxide material, channel dimensions), 
so desired characteristics can be easily determined and compared before embarking 
on an involved and time-consuming fabrication process. This also allows for 
prediction of ideal device and/or circuit behaviour, against which realised 
devices can be evaluated. 
Our second motivation is that our software has 
value as an educational tool. GFETs aren't available as off-the-shelf 
components in the same scale, quantities or price as conventional transistors, 
and so the ability to experiment with GFET behaviour, based on verified models 
is valuable for training and familiarising non-experts with these devices.

As part of this, we provide the ability to export models for use in SPICE simulators such as LTSpice. We have only implemented the model discussed in the article, but the typical format of an exported SPICE model is as follows:

\begin{lstlisting}
	* n1=Top Gate, n2=Drain, n3=Source, n4=Back Gate 
	.subckt GFETModel n1 n2 n3 n4 
	.params 
	+(Physical and geometric constants) 
	.func x={}
	...
	B1 n1 n3 I = (Drain Current Expression)
	.ends GFETModel 
\end{lstlisting}

We encourage others to contribute and expand on the software to maximise its utility,
either by expanding the library of available GFET models by translating others from
the literature, by expanding the functionality to better represent experiments one might
conduct using GFETs, or by submitting bug reports. 

What follows is a demonstrative description of model implementation in GFET Lab. 
Each model is a class which 
takes as arguments the contents of the user-defined sweep and device parameter 
boxes. Each model has the following methods: \texttt{init}, \texttt{fnIds}, \texttt{calculateTransferChars}, and 
\texttt{calculateIVChars}. The first initialises the model class with the relevant device 
and sweep parameters, the second performs the calculation of the drain current, the third simulates the device transfer characteristics, transconductance and transit frequency, and the fourth simulates the I-V characteristics. Additional functions, such as those to determine channel voltage or saturation velocity may also be implemented.  In principle, GFET Lab can be extended to model other 2D materials by implementing a suitable model, enhancing its value as a research tool.

\begin{lstlisting}
class GFETModel:
	
	def __init__(self, params, ivSweep, transSweep, eps):
	 		 self.ivVds = ivSweep["Vds"]
			 self.ivVtg = ivSweep["Vtg"]
			 self.ivVbg = ivSweep["Vbg"]
			 self.transVds = transSweep["Vds"]
			 self.transVtg = transSweep["Vtg"]
			 self.transVbg = transSweep["Vbg"]
			 ...
			 
			 (Model-dependent parameters, e.g., 
			   geometry, calculation of capaitances,
			   contace resistances...)
	 ...
	 
	 (Additional functions required by the model) 
	 ...
	 
	def fnIds(self, Vtg, Vbg, Vds):
		(Expressions required, e.g., to determine effective
		  channel length)
		return (Model expression of the drain current)
	
	def calculateTransferChars(self):
        Ids = []
		for Vds in self.transVds:
			Ids.append([self.fnIds(Vtg, self.transVbg, Vds) for Vds in self.transVtg])
		
		(Calculation of transconductance and transit frequency)
		
		return Ids, gm, fT, Vg
		
  def calculateIVChars(self):
		Ids = []
		for Vtg in self.ivVtg:
			Ids.append([self.fnIds(Vtg, self.ivVbg, Vds) for Vds in self.ivVds])
		return Ids
\end{lstlisting}
	%
%
\section{Conclusions}
\label{section:conclusions}
We have surveyed the GFET modelling literature, providing both a background and brief overview of the field, as well as discussing limitations of device modelling. Based on the survey, we identify three criteria for a suitable predictive device model. Such a model should be compact, SPICE-compatible, and require a minimal number of fitting parameters. Based on these criteria, we selected Jimenez et al.'s explicit drain-current GFET model and implemented it in a user-friendly software tool called \textit{GFET Lab}. We then simulate GFET characteristics based on the geometries and materials of real devices in the software and compare the simulations to empirically-measured characteristics of the devices to validate the model and software. The characteristics show good agreement with the real devices, being of the correct order of magnitude for the same voltage sweeps and bias conditions, with the main discrepancy being the on/off ratio. The reasonable agreement between simulation and experiment demonstrates GFET Lab's utility as both a research and educational tool. We also propose the notion of \textit{process parameters} for device modelling, where fitting parameters are determined as functions of the fabrication process, allowing for models to account for variations which arise as a result of the fabrication process, e.g., doping due to wet-transfer, or the more pronounced asymmetry between n-type and p-type conductance in CVD graphene compared to exfoliated graphene. We then demonstrate good agreement between the simulated characteristics of \textit{GFET Lab} and the exported SPICE models when simulated in LTSpice. Lastly, we provide a short tutorial to \textit{GFET Lab}, as well as detail how members of the community can contribute to the software to maximise its potential as a research tool.

%
%
%
%
%
%
%
%

	\vspace{0.25in}
	\section*{Acknowledgements}
	P. Stanley-Marbell is supported by an Alan Turing Institute award TU/B/000096 under EPSRC grant EP/N510129/1, by EPSRC grant EP/V047507/1, and by the UKRI Materials Made Smarter Research Centre (EPSRC grant EP/V061798/1). N.J. Tye acknowledges funding from EPSRC 
	grant EP/L016087/1. We also thank David Jimenez for his advice on model implementation, particularly in regards to considering intrinsic and extrinsic voltages, Giorgio Mallia for the photographs on the experimental setup in Figure~\ref{figure:ExperimentalSetup}, and Bilgesu Bilgin for fabricating GFET1 which we characterise in this article.
	
	\section*{Code Availability}
	The full source code for GFET Lab is available at: https://github.com/physical-computation/gfet-simulator.
	
	\bibliographystyle{abbrv}
	\bibliography{working-document}
	
\end{document}